\documentclass[aps,prl,reprint]{revtex4-2}
\usepackage{blindtext}
\usepackage{graphicx}
\usepackage{subcaption}
\usepackage{xcolor}
\usepackage[fleqn]{amsmath}
\usepackage[font=small,skip=3pt]{caption}
\usepackage{makecell}
\usepackage{hyperref} 
\usepackage{cleveref}
\usepackage{url}
\usepackage{multirow}
\usepackage{listings}
\begin{document}
\title{HIPED: Machine Learning Framework for Spherical Tokamak Pedestal Prediction and Optimization}
\author{J. F. Parisi$^{1}$}
\email{jparisi@pppl.gov}
\author{J. G. Clark$^1$}
\author{J. W. Berkery$^1$}
\author{C. Bowman$^2$}
\author{C. J. Fitzpatrick$^{3}$}
\author{S. M. Kaye$^1$}
\author{M. Lampert$^1$} 
\affiliation{$^1$Princeton Plasma Physics Laboratory, Princeton University, Princeton, NJ, USA}
\affiliation{$^2$United Kingdom Atomic Energy Authority, Culham Campus, Abingdon, Oxfordshire, OX14 3DB, United Kingdom}
\affiliation{$^3$Harvey Mudd College, 301 Platt Blvd, Claremont, CA 91711}

\begin{abstract}
We introduce a Machine Learning framework, HIPED (HeIght and width Predictor for Edge Dynamics), for predicting and optimizing pedestal and core performance in spherical tokamak plasmas. Trained on pedestal and core datasets from the third MAST-U campaign, HIPED provides accurate estimates of pedestal height and width. The results reveal notable differences compared with conventional aspect-ratio studies; for instance, a simple power-law relation between pedestal width and height has very low accuracy. Instead, additional parameters such as normalized plasma pressure, elongation, and Greenwald fraction significantly improve accuracy. HIPED can also be trained only on `control room parameters' to inform experimentalists of which controllable parameters to adjust for improving core-integrated performance. The framework further includes a multi-objective optimization scheme that helps guide experimental planning and optimization. We find Pareto-optimal discharges with respect to various features, including distance from edge-localized modes and normalized plasma pressure, track their parameter trajectories over time, and identify the control room parameters required for these Pareto-optimal discharges. This provides a framework for systematically optimizing core and edge performance according to different experimental priorities.
\end{abstract}

\maketitle

\setlength{\parskip}{0mm}
\setlength{\textfloatsep}{5pt}

\setlength{\belowdisplayskip}{6pt} \setlength{\belowdisplayshortskip}{6pt}
\setlength{\abovedisplayskip}{6pt} \setlength{\abovedisplayshortskip}{6pt}

\section{Introduction}

Magnetic confinement fusion power plants rely on achieving high fusion power density in a hot, dense plasma core alongside a cooler plasma edge. In tokamaks, one efficient approach to establishing such conditions is through H-mode, a high-confinement regime distinguished by steep plasma edge profiles \cite{Wagner1982,Kaye1984,Ryter1996}. The pedestal—a region near the plasma edge characterized by sharp gradients in density and temperature—plays a critical role in determining global confinement quality, fusion power output, and overall tokamak performance \cite{Shimada2007,Wenninger2015,Sorbom2015,Snyder2019,Creely2020,Rodriguez-Fernandez2022,Osborne2023}. Accurate prediction of pedestal profiles is thus essential for optimizing the operation of fusion power plants in H-mode. For instance, simulations for the SPARC experiment indicate that a 50\% reduction in the temperature pedestal could decrease fusion gain ($Q$) by up to two-thirds \cite{Creely2020,Hughes2020}. However, reliable pedestal prediction remains challenging due to the complexity of the edge region, where transport processes, heat, momentum and particle sources, neoclassical physics, and magnetohydrodynamic (MHD) stability are intricately coupled \cite{Kotschenreuther2019,Mordijck2020,Groebner_2023}.

Pedestal modeling has often relied on either empirical scaling laws or computationally expensive physics-based models \cite{Groebner2002,Snyder2009,Snyder2011,Merle2017,Saarelma2019,Saarelma_2024,Parisi_2024}. Here, we introduce a machine learning (ML) framework called HIPED (HeIght and width Predictor for Edge Dynamics), which predicts pedestal parameters in spherical tokamaks and facilitates multi-objective optimization. HIPED leverages an extensive dataset from the MAST-U device \cite{Harrison2024} to train models that accurately predict pedestal height and width, yielding new insight into pedestal formation at low aspect ratio. Automated pedestal fits are performed on this database using a Bayesian multi-diagnostic inference
system \cite{Bowman2020,Greenhouse2025,Clark2025}. Additionally, we show how restricting the training to `control room parameters’ illuminates which levers experimentalists can directly adjust to enhance overall plasma performance. HIPED also employs Pareto optimization, making it possible to identify regimes that optimize multiple objectives simultaneously (for example, high confinement while avoiding edge-localized modes (ELMs) \cite{Kirk2004,Osborne2015}).

Machine learning has become increasingly important in plasma physics. It has been used for disruption prediction \cite{Rea2019,Vega2022,Sabbagh2023,Gambrioli_2025}, turbulence acceleration \cite{Ma2020,Van2020}, real-time control \cite{Degrave2022}, equilibrium reconstruction \cite{Felici2011,Lao2022,Candido2023}, plasma heating \cite{Wallace2022,Sanchez2024}, stability assessment \cite{Piccione2020,Piccione2022} and profile prediction \cite{Boyer2021,Abbate2021,Dubbioso2023}, among other topics \cite{Smith2013,Meneghini2021,Kit2023,vanLeeuwen_2025,Pavone2023,Landreman2025}. Past ML-based pedestal studies on conventional aspect ratio tokamaks (e.g. DIII-D, JET) have shown strong promise \cite{KatesHarbeck2019,Gillgren_2022}, but fewer studies have focused on pedestals in low-aspect-ratio devices such as MAST \cite{Smith2022,Harrison2024} or NSTX \cite{Smith2016,Berkery_2024}, even though pedestal physics can be markedly different at low aspect ratio.

A consensus on pedestal width-height scalings has yet to emerge \cite{Viezzer2023}. Earlier studies found that the pedestal width-height scaling follows $\Delta_{\mathrm{ped}} \sim \sqrt{\beta_{\theta,\mathrm{ped}}}$ for ELMy H-modes in the conventional aspect ratio DIII-D tokamak \cite{Groebner1998}. These findings were corroborated by the EPED model and a database study \cite{Snyder2009b} that demonstrated excellent agreement with 4,122 DIII-D time slices. However, a study on JET—also a conventional aspect ratio tokamak—found that the pedestal width did not follow the EPED dependence $\Delta_{\mathrm{ped}} = 0.076 \sqrt{\beta_{\theta,\mathrm{ped}}}$ \cite{Frassinetti2020}, with a spread in the coefficient of a factor of order unity. A separate ML model, PENN, was trained using a JET pedestal database and achieved high $R^2$ values for pedestal density and temperature prediction ($R^2 = 0.96, 0.91$, respectively) \cite{Gillgren_2022}. A 2003 ITPA database study across seven machines (AUG, C-Mod, DIII-D, JET-C, JFT-2M, JT-60U, and MAST) fitted a different definition of pedestal height with a root-mean-square error of approximately 20\% \cite{Cordey_2003}.

Building on previous MAST/MAST-U pedestal investigations \cite{Smith2022,Clark2025}, we find that standard scaling laws like $\Delta_{\mathrm{ped}} \sim \sqrt{\beta_{\theta,\mathrm{ped}}}$ \cite{Snyder2009} do not hold as neatly for low-aspect-ratio data. Instead, including parameters like the total normalized pressure, plasma elongation, and Greenwald fraction significantly improve predictive accuracy. We also use Pareto optimization to map out the tradeoffs among multiple objectives, such as distance in time to ELMs, Greenwald fraction \cite{Greenwald1988,Berkery2023}, line-averaged electron density, and normalized plasma pressure. This approach highlights which operating parameters facilitate the best performance tradeoffs in experiments.

This paper is organized as follows. \Cref{sec:pedestalcharacteristics} reviews pedestal definitions. \Cref{sec:pedfit_methods} describes the fits that extract pedestal parameters from MAST-U. \Cref{sec:MASTUdatabase} details the MAST-U dataset used and applies linear regression to show how conventional scaling laws are inadequate for spherical tokamaks. \Cref{sec:MLmodel} introduces the Random Forest ML approach and shows how restricting training to control room parameters can still yield robust predictions of pedestal density and temperature. \Cref{sec:Pareto_optimization} describes HIPED’s multi-objective optimization and applies it to identify Pareto-optimal discharges. \Cref{sec:discussion} summarizes the key findings and future directions. Additional information is given in the Appendices.

\section{Pedestal Parameters}
\label{sec:pedestalcharacteristics}

\begin{figure}[bt!]
    \centering
    \begin{subfigure}[t]{0.49\textwidth}
    \centering
    \includegraphics[width=\textwidth]{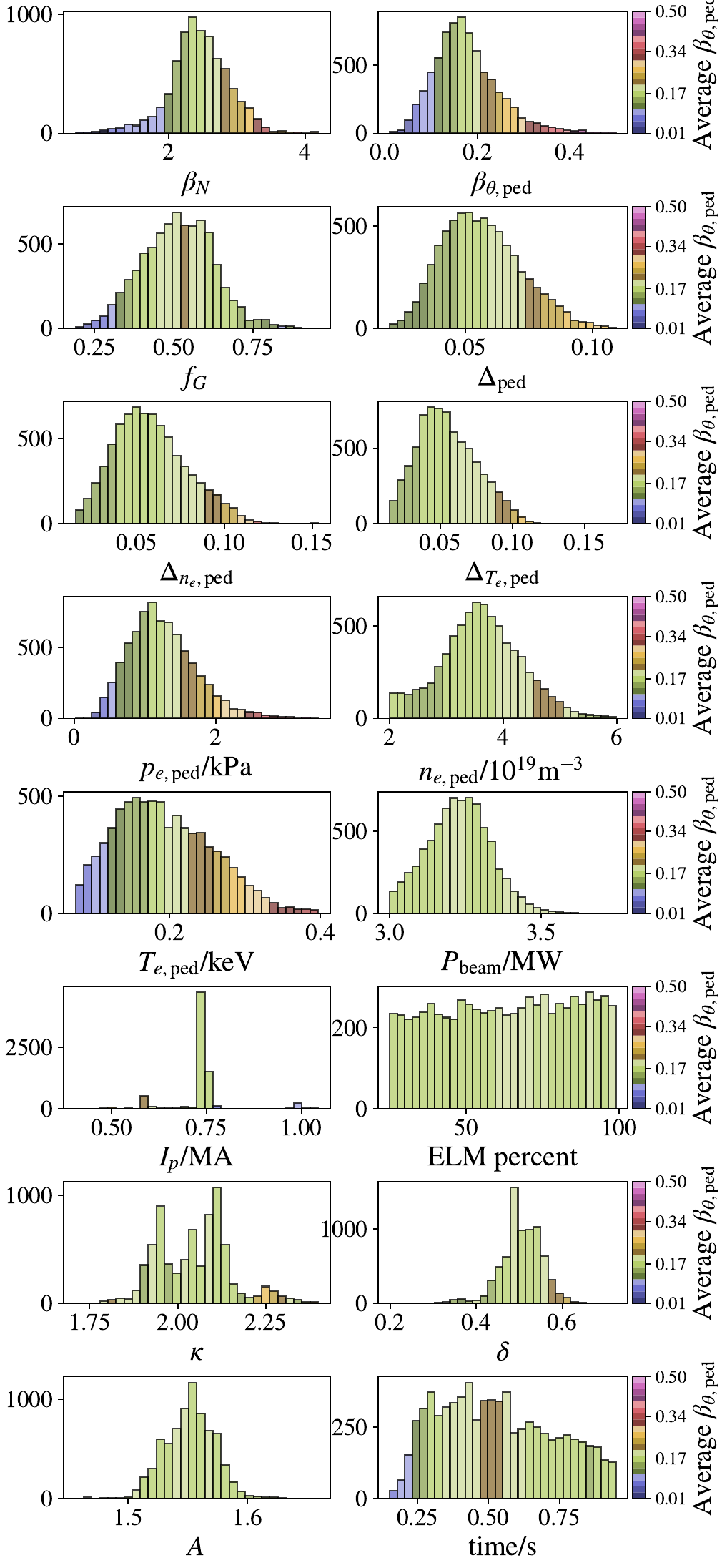}
    \end{subfigure}
    \caption{Histograms of representative pedestal and global parameters for the `H-mode all' category in the MAST-U database. The color scale shows the mean $\beta_{\theta,\mathrm{ped}}$ value.}
    \label{fig:MASTU_database_histograms}
\end{figure}

In this section, we introduce the pedestal parameters used in this work. We adopt the EPED framework \cite{Snyder2009} for parameter definitions, which motivates studying pedestal width $\Delta_{\mathrm{ped}}$ in relation to normalized pedestal pressure $\beta_{\theta,\mathrm{ped}}$. The edge electron density and temperature profiles, $n_e(\psi_N)$ and $T_e(\psi_N)$, are often fitted to hyperbolic tangent (tanh) forms for the electron density
\begin{equation}
\begin{aligned}
& n_e(\psi_N) = n_{e,\mathrm{core}} \mathrm{H}\bigl[\psi_{\mathrm{ped,n_e}}-\psi_N\bigr]\,(1-\psi_N^{\alpha_{n_1}})^{\alpha_{n_2}} \\
& \quad +\, n_{e0}\bigl(t_1 - \tanh\bigl(\tfrac{\psi_N-\psi_{\mathrm{mid,n_e}}}{\,\Delta_{n_e,\mathrm{ped} }/2}\bigr)\bigr) + n_{\mathrm{e,off}},
\end{aligned}
\label{eq:1}
\end{equation}
and electron temperature
\begin{equation}
\begin{aligned}
& T_e(\psi_N) = T_{e,\mathrm{core}} \mathrm{H}\bigl[\psi_{\mathrm{ped,T_e}}-\psi_N\bigr]\,(1-\psi_N^{\alpha_{T_1}})^{\alpha_{T_2}} \\
& \quad +\, T_{e0}\bigl(t_1 - \tanh\bigl(\tfrac{\psi_N-\psi_{\mathrm{mid,T_e}}}{\,\Delta_{T_e,\mathrm{ped} }/2}\bigr)\bigr) + T_{\mathrm{e,off}},
\end{aligned}
\label{eq:2}
\end{equation}
where $\psi_N$ is the poloidal flux normalized to 0 on the magnetic axis at 1 at the last-closed flux surface, H is a step function, $t_1=\tanh(1)$, and $\Delta_{n_e,\mathrm{ped} }$, $\Delta_{T_e,\mathrm{ped} }$ are the density and temperature pedestal widths in $\psi_{N}$. The $\alpha$ quantities are exponent fit parameters, $n_{e,\mathrm{core}}$, $T_{e,\mathrm{core}}$, $n_{e0}$, and $T_{e0}$ are constants, and $n_{e,\mathrm{off}}$ and $T_{e,\mathrm{off}}$ are evaluated at $\psi_N = \psi_{\mathrm{ped,n_e}} + \Delta_{n_e,\mathrm{ped} }$ and $\psi_N = \psi_{\mathrm{ped,T_e}} + \Delta_{T_e,\mathrm{ped} }$. The density and temperature pedestal heights are defined as
\begin{equation}
n_{e,\mathrm{ped}} = n\bigl(\psi_{\mathrm{ped,n_e}}\bigr), \quad T_{e,\mathrm{ped}} = T\bigl(\psi_{\mathrm{ped,T_e}}\bigr).
\end{equation}
We define the total pedestal width and the ``top'' location via
\begin{equation}
\Delta_{\mathrm{ped}} \equiv \frac{\Delta_{n_e,\mathrm{ped} } + \Delta_{T_e,\mathrm{ped} }}{2},
\quad
\psi_{\mathrm{ped}} \equiv \psi_{\mathrm{mid}} - \frac{\Delta_{\mathrm{ped}}}{2},
\end{equation}
where $\psi_{\mathrm{mid}} = (\psi_{\mathrm{mid},n_e}+\psi_{\mathrm{mid},T_e})/2$. The pedestal height is often characterized by $\beta_{\theta,\mathrm{ped}}$:
\begin{equation}
    \beta_{\theta, \mathrm{ped}} \equiv \frac{2 \mu_0 p_{\mathrm{ped}}}{\overline{B}_{\mathrm{pol}}^2},
    \quad
    p_{\mathrm{ped}} \equiv 2\,p_e(\psi_{\mathrm{ped}}),
\end{equation}
where $p_{e}=n_{e}T_{e}$, $\overline{B}_{\mathrm{pol}} = \mu_0 I_p/l$, $\mu_0$ is the magnetic permittivity of free space, and $l$ is the last-closed-flux-surface circumference. Throughout this work, the quantities $\beta_{\theta, \mathrm{ped} }$, $\Delta_\mathrm{ped}$, $n_{e,\mathrm{ped} }$, $T_{e,\mathrm{ped} }$, $\Delta_{n_e,\mathrm{ped} }$, and $\Delta_{T_e,\mathrm{ped}}$ will be frequently used.

\section{Pedestal Fit Methodology}
\label{sec:pedfit_methods}

In this work, we fit the electron density and temperature profiles to tanh-based functions [\Cref{eq:1,eq:2}] following standard pedestal-fitting approaches. We then apply a series of quality checks:
\begin{enumerate}
    \item The fit must converge and produce physically valid parameters (e.g.\ positive pedestal width).
    \item We discard noise-induced outliers based on residual thresholds.
    \item We only include fits taken in a suitable fraction of the ELM cycle so that each point genuinely represents an H-mode interval.
\end{enumerate}
This yields a set of valid density and temperature pedestal parameters. See Ref.~\cite{Clark2025} for more details on the pedestal-fitting routines and \cite{Bowman2020,Greenhouse2025} for more details on the Bayesian multi-diagnostic inference system.

\section{MAST-U Pedestal Database}
\label{sec:MASTUdatabase}

\begin{figure*}[tb]
    \centering
    \begin{subfigure}[t]{0.32\textwidth}
    \includegraphics[width=1.0\textwidth]{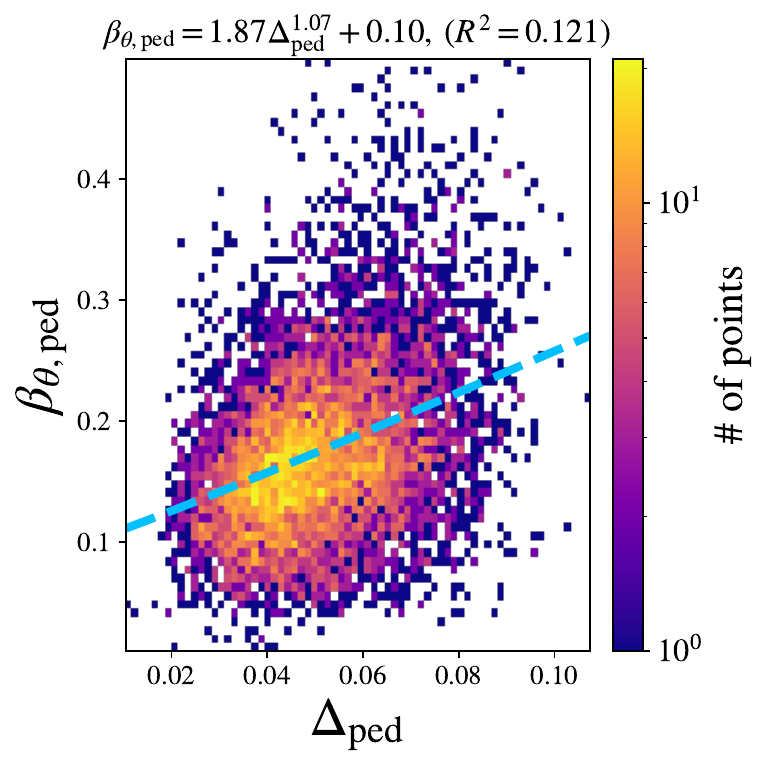}
    \caption{H-mode all}
    \end{subfigure}
    \begin{subfigure}[t]{0.32\textwidth}
    \includegraphics[width=1.0\textwidth]{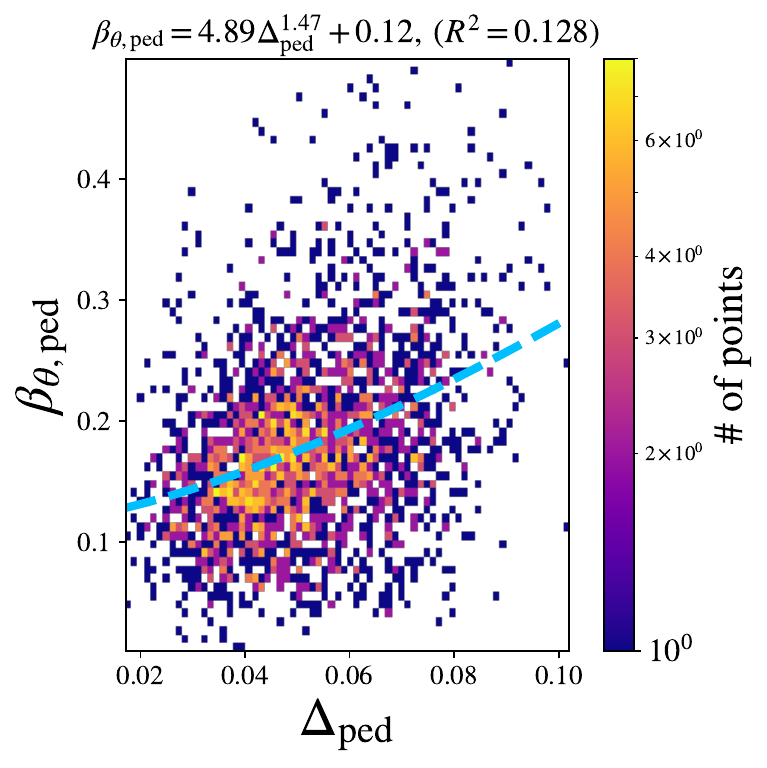}
    \caption{pre-ELM}
    \end{subfigure}
    \begin{subfigure}[t]{0.32\textwidth}
    \includegraphics[width=1.0\textwidth]{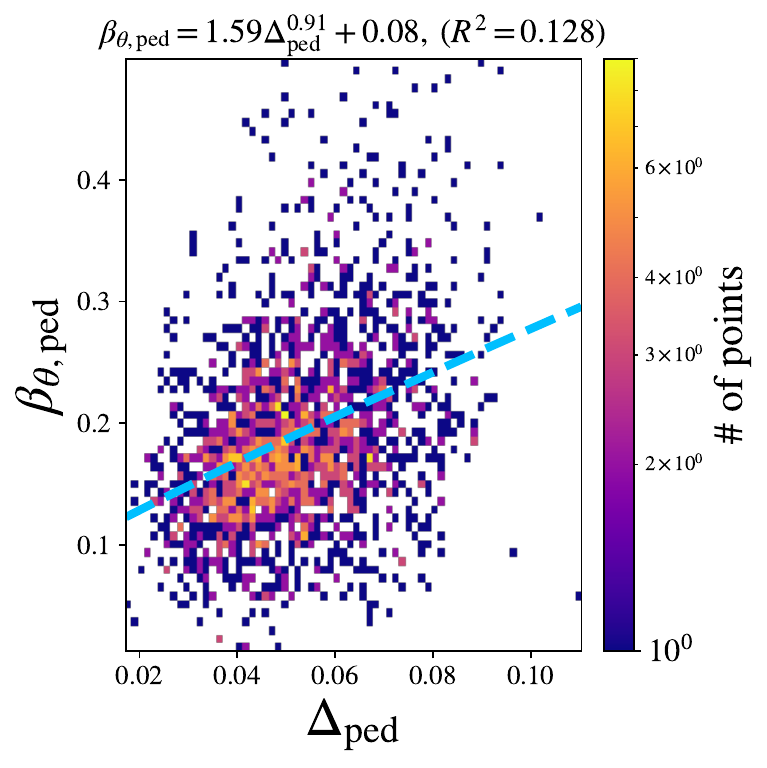}
    \caption{ELM-free}
    \end{subfigure}
    \begin{subfigure}[t]{0.32\textwidth}
    \includegraphics[width=1.0\textwidth]{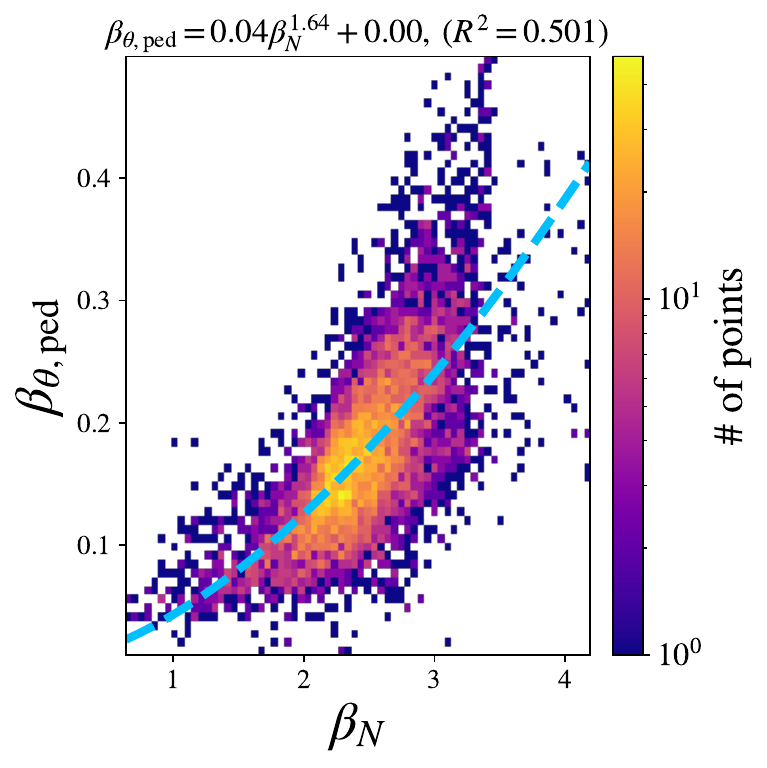}
    \caption{H-mode all}
    \end{subfigure}
    \begin{subfigure}[t]{0.32\textwidth}
    \includegraphics[width=1.0\textwidth]{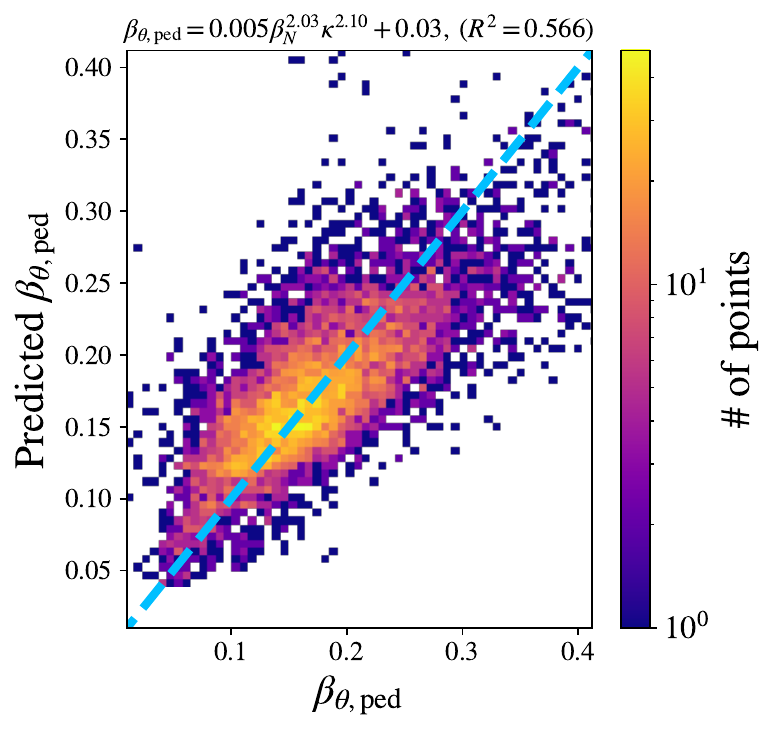}
    \caption{H-mode all}
    \end{subfigure}
    \begin{subfigure}[t]{0.32\textwidth}
    \includegraphics[width=1.0\textwidth]{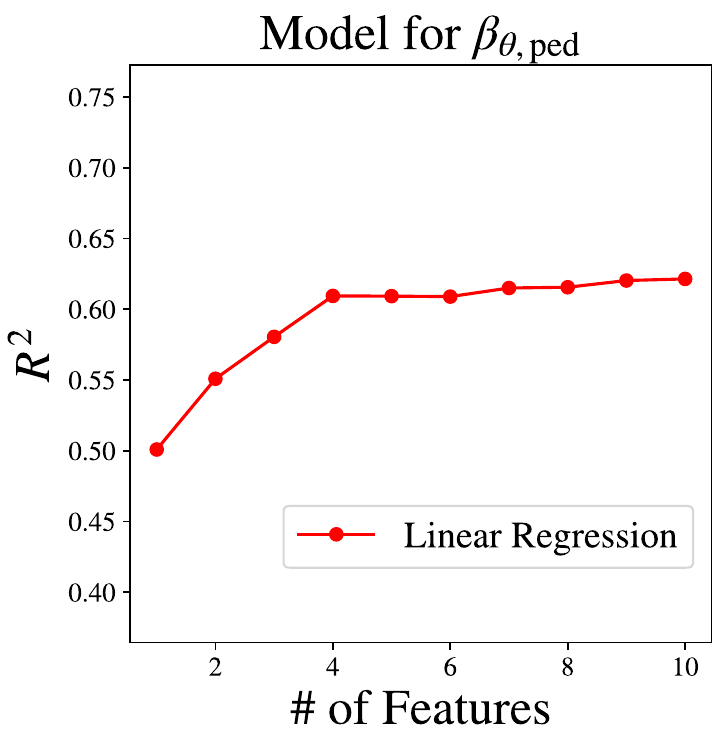}
    \caption{H-mode all}
    \end{subfigure}
    \caption{(a)-(c)~Scatter plots of $\Delta_{\mathrm{ped}}$ versus $\beta_{\theta,\mathrm{ped}}$ for the three MAST-U filter categories. The dashed line is a fit $\Delta_{\mathrm{ped}}=c_1(\beta_{\theta,\mathrm{ped}})^{c_2}+c_3$, which yields a low $R^2<0.13$ in each case. (d)-(e)~Least-squares regression of $\beta_{\theta,\mathrm{ped}}$. (d)~Using only $\beta_{N}$. (e)~Using $\beta_{N}$ and $\kappa$. (f)~$R^2$ versus number of features (H-mode all).}
    \label{fig:MASTUdatabase_full}
\end{figure*}

\begin{figure*}[]
    \centering
    \begin{subfigure}[t]{0.32\textwidth}
    \centering
    \includegraphics[width=\textwidth]{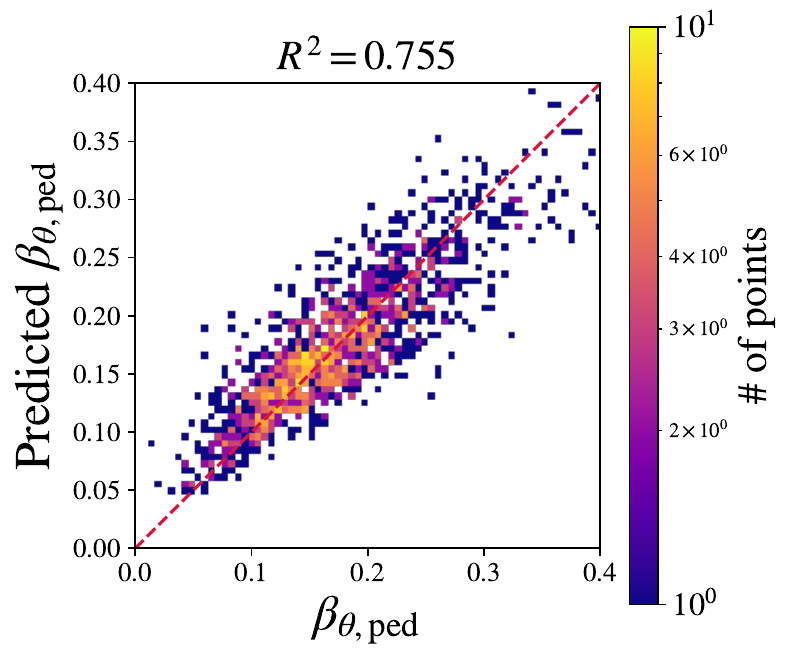}
    \caption{H-mode all}
    \end{subfigure}
    \begin{subfigure}[t]{0.32\textwidth}
    \centering
    \includegraphics[width=\textwidth]{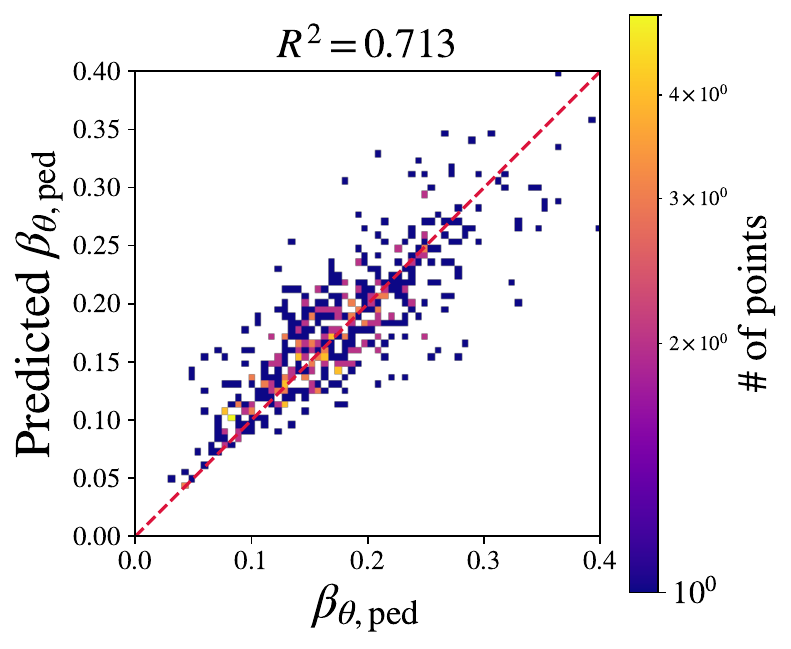}
    \caption{pre-ELM}
    \end{subfigure}
    \begin{subfigure}[t]{0.32\textwidth}
    \centering
    \includegraphics[width=\textwidth]{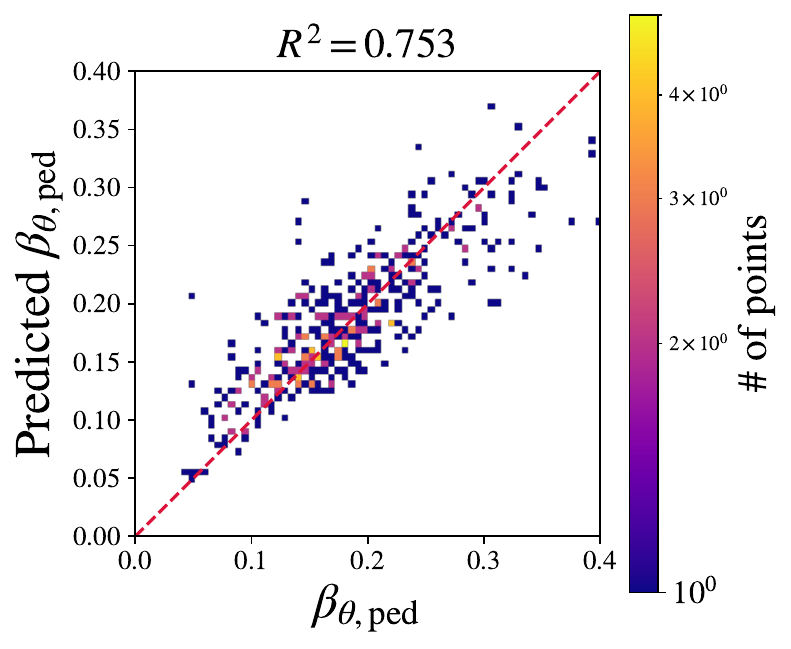}
    \caption{ELM-free}
    \end{subfigure}
    \begin{subfigure}[t]{0.32\textwidth}
    \centering
    \includegraphics[width=\textwidth]{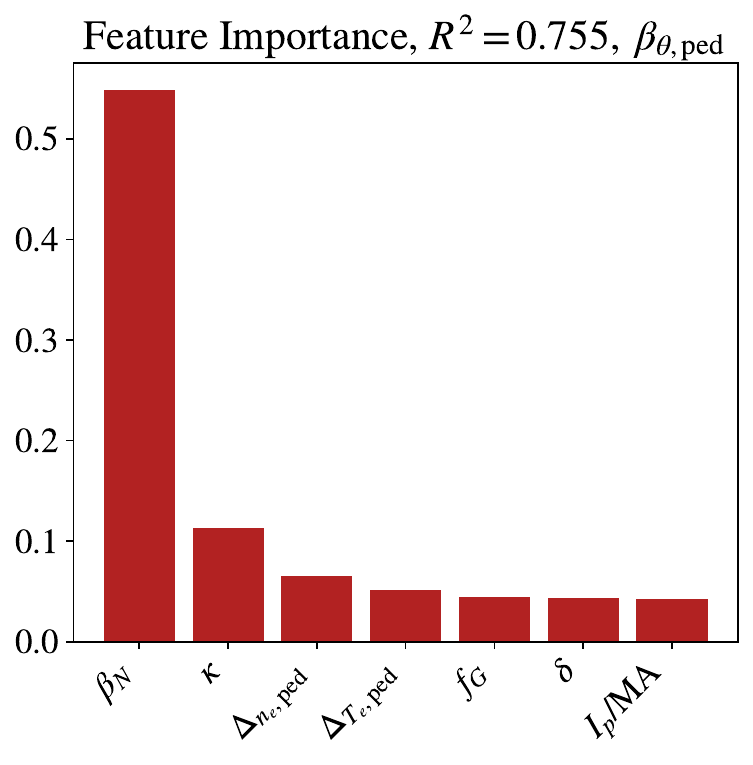}
    \caption{H-mode all}
    \end{subfigure}
    \begin{subfigure}[t]{0.32\textwidth}
    \centering
    \includegraphics[width=\textwidth]{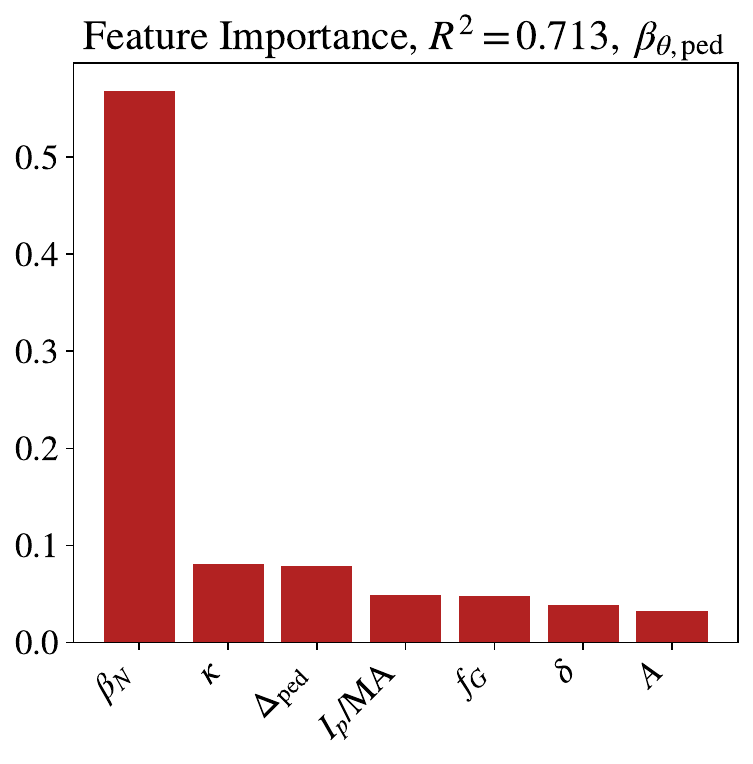}
    \caption{pre-ELM}
    \end{subfigure}
    \begin{subfigure}[t]{0.32\textwidth}
    \centering
    \includegraphics[width=\textwidth]{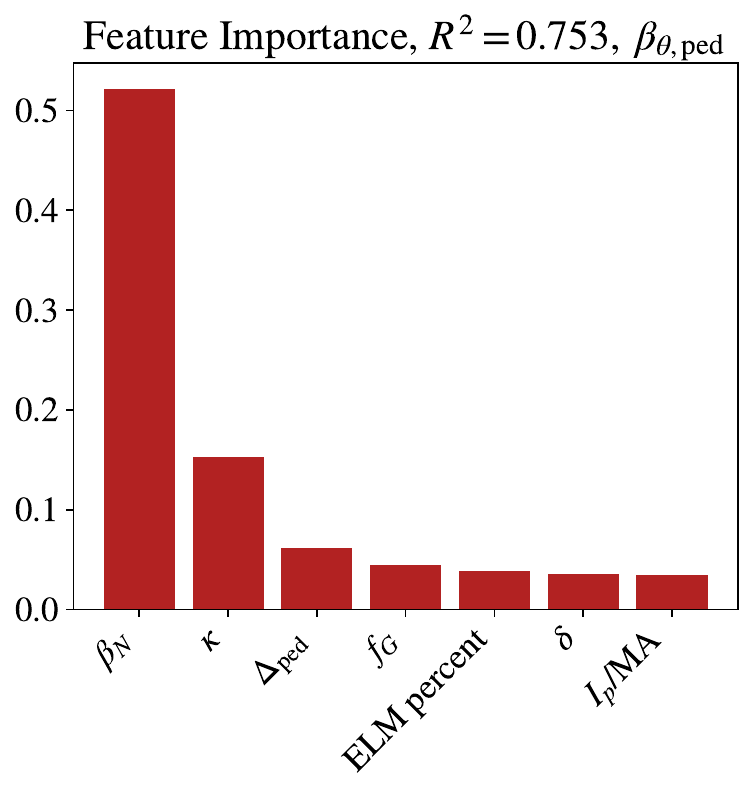}
    \caption{ELM-free}
    \end{subfigure}
    \caption{Random Forest predictions of $\beta_{\theta,\mathrm{ped}}$ in the three filter scenarios. Top row: target (true) distributions. Bottom row: predicted vs.\ actual (test set). $R^2\approx 0.71$--0.76.}
    \label{fig:betapedtheta_RF}
\end{figure*}

\begin{table}[b!]
\centering
\begin{tabular}{|c|c|c|c|c|c|c|}
\hline
\multirow{2}{*}{\textbf{Quantity}} & 
\multicolumn{2}{|c|}{\textbf{H-mode}} & 
\multicolumn{2}{|c|}{\textbf{pre-ELM}} & 
\multicolumn{2}{|c|}{\textbf{ELM-free}} \\ 
\cline{2-7}
& Min & Max & Min & Max & Min & Max \\
\hline
ELM percent (\%) & 25 & 99 & 75 & 99 & - & - \\
\hline
$\beta_{\theta,\mathrm{ped}}$ & 0.03 & 0.50 & 0.03 & 0.50 & 0.03 & 0.50 \\
\hline
$n_{e,\mathrm{ped}}$ /$10^{19}$m$^3$ & 2 & 6 & 2 & 6 & 2 & 6 \\
\hline
$T_{e,\mathrm{ped}}$ /(eV) & 75 & 400 & 75 & 400 & 75 & 400 \\
\hline
$\Delta_{\mathrm{ped}}$ & 0.02 & 0.11 & 0.02 & 0.11 & 0.02 & 0.11 \\
\hline
$\Delta_{n_e,\mathrm{ped}}$ & 0.01 & 0.20 & 0.01 & 0.20 & 0.01 & 0.20 \\
\hline
$\Delta_{T_e,\mathrm{ped}}$ & 0.01 & 0.20 & 0.01 & 0.20 & 0.01 & 0.20 \\
\hline
$P_{\mathrm{beam}}$/MW & 3 & $\infty$ & 3 & $\infty$ & 3 & $\infty$ \\
\hline
$\kappa$ & 1.7 & 2.4 & 1.7 & 2.4 & 1.7 & 2.4 \\
\hline
\end{tabular}
\caption{Filters applied to the MAST-U pedestal database. Only the ELM percent quantity differs among the three categories.}
\label{table:MASTU_three_filters}
\end{table}

Our data comes from the third MAST-U campaign, initially containing 711 shots and 65,393 equilibria spanning a range of shaping parameters, neutral beam powers, and ELM behaviors. For this study, we restrict attention to H-mode discharges by applying filters to form three main categories:
\begin{enumerate}
\item \emph{H-mode all}: Data from $25\%$ to $99\%$ of the ELM cycle (7481 points).
\item \emph{H-mode pre-ELM}: Data from $75\%$ to $99\%$ of the ELM cycle (2582 points).
\item \emph{H-mode ELM-free}: At least 0.05 seconds before and after an ELM (2117 points).
\end{enumerate}
These filters, summarized in \Cref{table:MASTU_three_filters}, exclude unphysical outliers (e.g.\ pedestal density below $10^{19}\,\mathrm{m}^{-3}$). The ELM-free filter for the third category calculates the time since and until an ELM for all timeslices. All timeslices less than 0.05 seconds before or after an ELM are excluded.

\begin{figure*}[!tb]
    \centering    
    \begin{subfigure}[t]{0.75\textwidth}
    \centering
    \includegraphics[width=\textwidth]{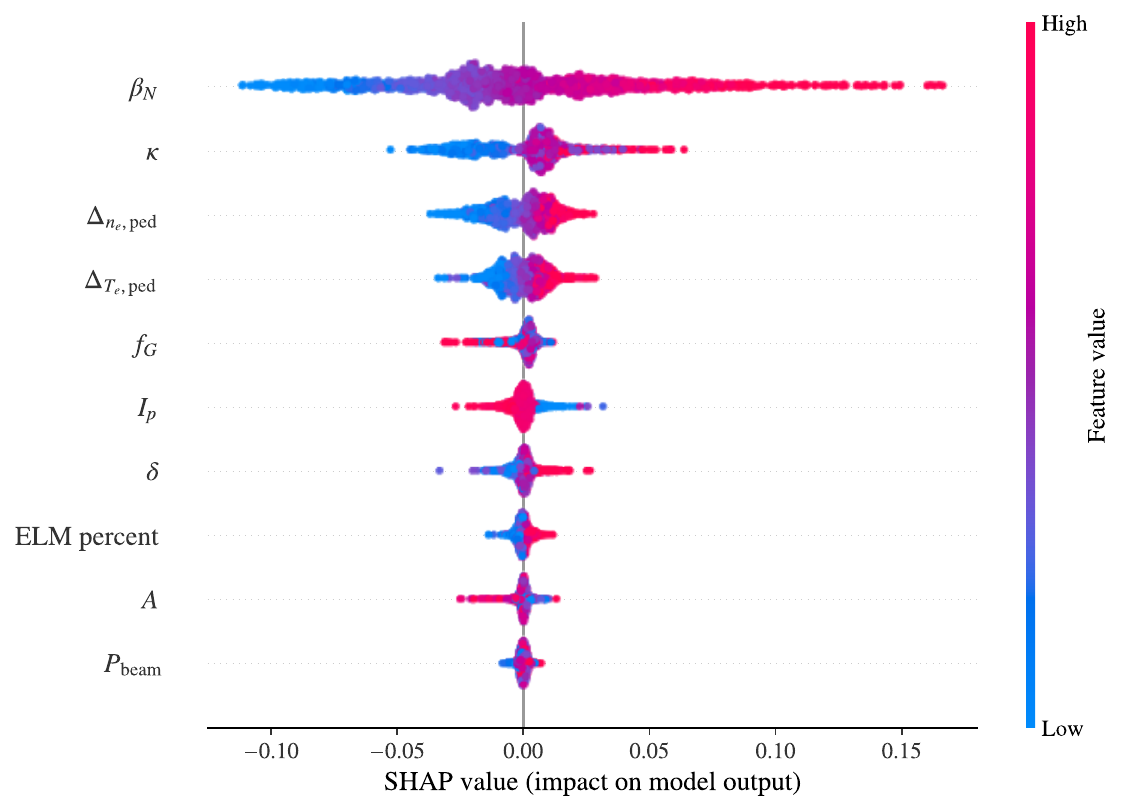}
    \end{subfigure}
    \caption{SHAP analysis of the H-mode all RF model for $\beta_{\theta,\mathrm{ped}}$. A positive (negative) SHAP value indicates that feature increases (decreases) the predicted pedestal height.}
    \label{fig:shap_RF_Model_betaped}
\end{figure*}

\Cref{fig:MASTU_database_histograms} shows histograms of representative parameters for the `H-mode all' category. The color scale highlights $\beta_{\theta,\mathrm{ped}}$. Here, $\beta_N = ( 2\,\mu_0\,\langle p\rangle / B_{T,0}^2 ) (a B_{T,0})/I_p$ is the normalized plasma pressure where $\langle p \rangle $ is the volume-averaged pressure, $B_{T,0}$ is the toroidal field at the magnetic field, $a$ is the minor radius, and $I_p$ is the plasma current. The quantity $f_G$ is the Greenwald density fraction \cite{Greenwald1988}, $P_\mathrm{beam}$ is the total neutral beam bower, ELM percent is the percent of the ELM cycle that a particular timeslice lies at, $\kappa$ is the plasma elongation, $\delta$ is the plasma triangularity, and $A$ is the plasma aspect ratio.

\Cref{fig:MASTU_database_histograms} shows that $\beta_{\theta,\mathrm{ped}}$ has a strong correlation with the total normalized pressure $\beta_N$ but a weaker correlation with $\Delta_{\mathrm{ped}}$. Indeed, studies on conventional aspect-ratio devices often report $\Delta_{\mathrm{ped}}\propto\sqrt{\beta_{\theta,\mathrm{ped}}}$ \cite{Groebner1998,Snyder2009b}, but \Cref{fig:MASTUdatabase_full}(a)-(c) suggest this simple scaling does not apply well here. In each of the three filter categories, a fit of the form $\Delta_{\mathrm{ped}}=c_1(\beta_{\theta,\mathrm{ped}})^{c_2}+c_3$ -- where $c_1$, $c_2$, and $c_3$ are constants -- yields $R^2<0.13$, implying that pedestal width depends on more than just $\beta_{\theta,\mathrm{ped}}$ at low aspect ratio.

Fitting different variables to $\beta_{\theta,\mathrm{ped}}$ substantially improves accuracy. \Cref{fig:MASTUdatabase_full}(d) shows $\beta_{\theta,\mathrm{ped}}$ versus $\beta_{N}$ for the H-mode all database; a power-law fit has $R^2=0.50$, which is a marked improvement. Adding elongation $\kappa$ further boosts $R^2$ to 0.57 [\Cref{fig:MASTUdatabase_full}(e)], but incorporating additional features (those from \Cref{fig:MASTU_database_histograms}) saturates around $R^2=0.62$ [\Cref{fig:MASTUdatabase_full}(f)], indicating that simple linear or power-law regressions remain insufficient. This motivates us to use a more flexible machine learning approach.

\section{Machine Learning Model}
\label{sec:MLmodel}

We employ a Random Forest (RF) to predict pedestal parameters such as $\beta_{\theta,\mathrm{ped}}$, $n_{e,\mathrm{ped}}$, and $T_{e,\mathrm{ped}}$ from a broader set of engineering and physics parameters. An RF model trains an ensemble of decision trees on random data subsets and averages their predictions \cite{Breiman2001,Liaw2002}, typically resulting in robust performance with limited hyperparameter tuning. RFs can capture nonlinear interactions in high-dimensional datasets, and feature-importance analyses allow for interpretability.

\subsection{Data Input and Preparation}

We extract a broad set of potential input parameters, including shaping and geometry ($\kappa,\delta,A$), normalized pressure ($\beta_N$), Greenwald fraction ($f_G$), and neutral beam power ($P_{\mathrm{beam}}$). MAST-U routinely runs with on-axis (``south'') and off-axis (``southwest'') neutral beam injection \cite{Rivero2024}. In this work, we will routinely distinguish between the two beams. Their beam powers satisfy $P_{\mathrm{beam}} = P_{\mathrm{beam,ss}} + P_{\mathrm{beam,sw}}$. Features with Pearson correlation $>0.85$ are removed to reduce overfitting. The remaining data is randomly split: 30\% for training (with cross-validation for hyperparameter selection) and 70\% for testing. \Cref{tab:tab0}, shown in the appendix, lists the key input parameters for the MAST-U database used in this work.

\subsection{\texorpdfstring{$\beta_{\theta,\mathrm{ped}}$}{Lg} Model}

\begin{figure*}[!tb]
    \centering    
    \begin{subfigure}[t]{0.4\textwidth}
    \centering
    \includegraphics[width=\textwidth]{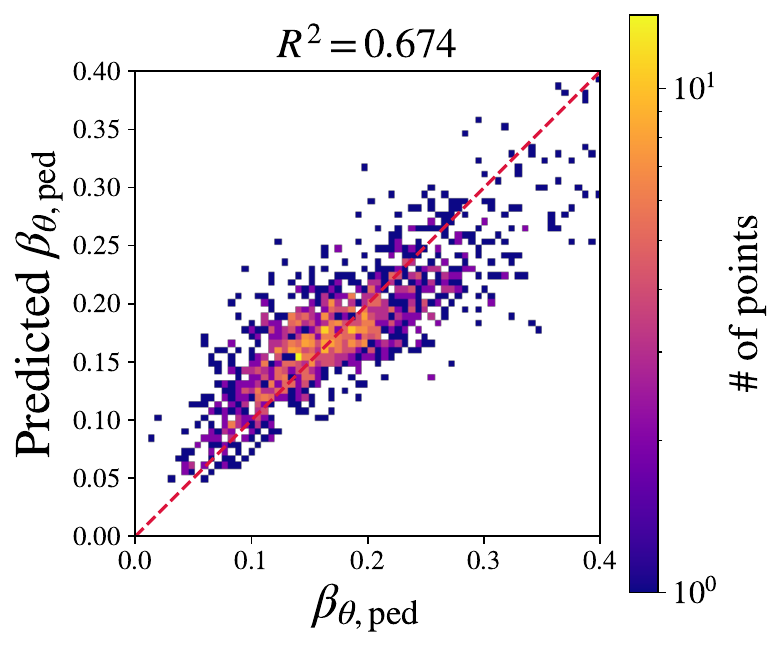}
    \caption{}
    \end{subfigure}
    \begin{subfigure}[t]{0.34\textwidth}
    \centering
    \includegraphics[width=\textwidth]{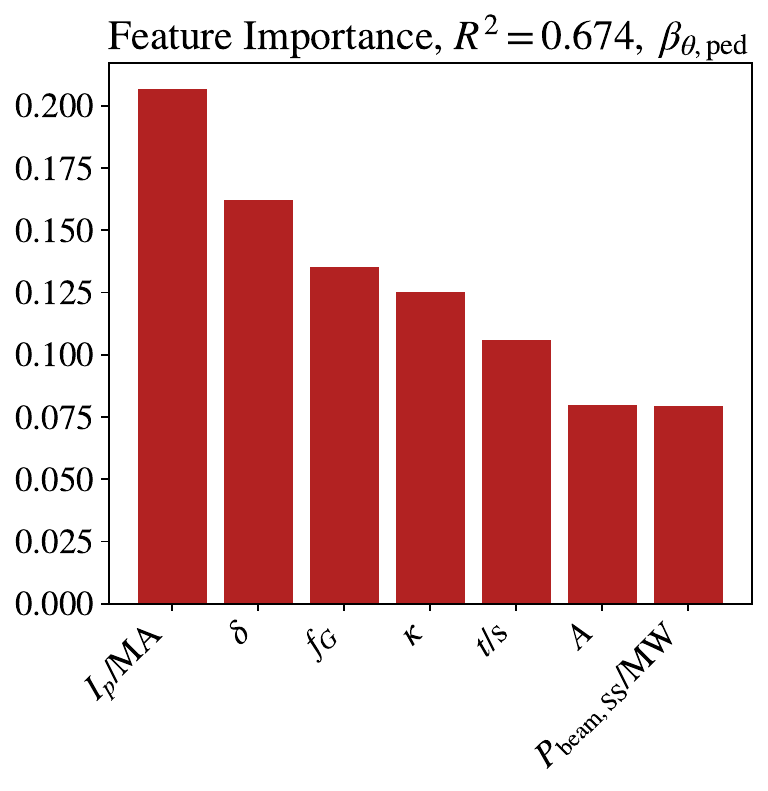}
    \caption{}
    \end{subfigure}
    \begin{subfigure}[t]{0.75\textwidth}
    \centering
    \includegraphics[width=\textwidth]{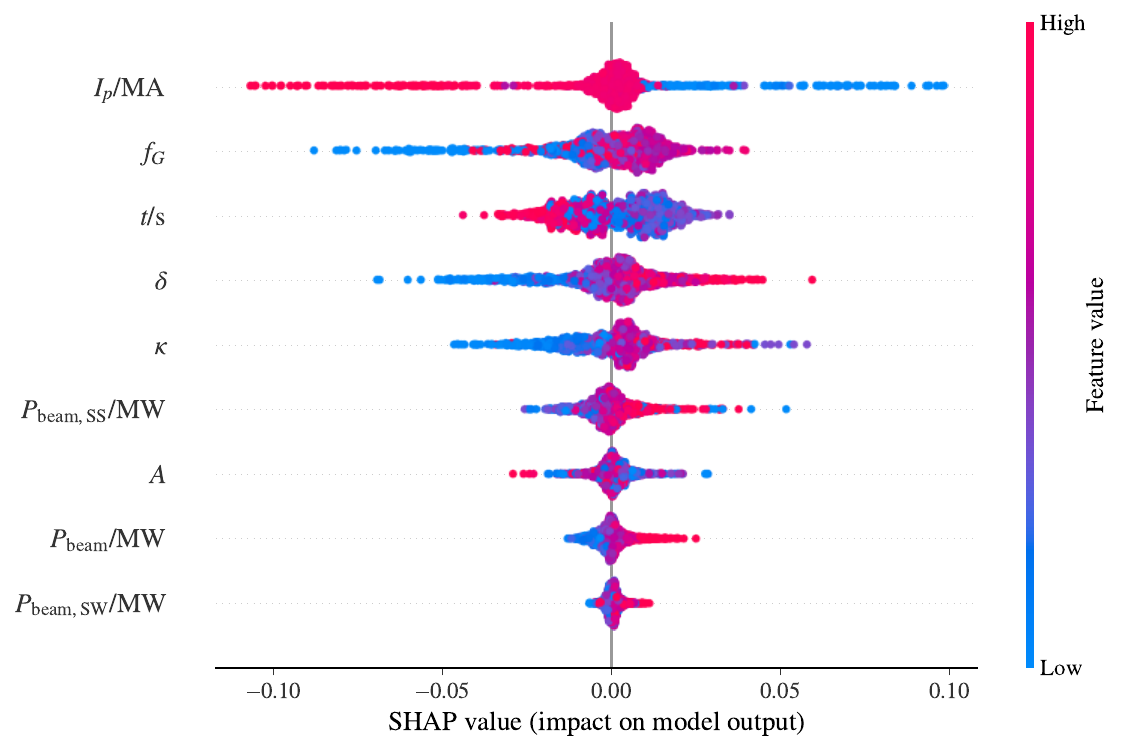}
    \caption{}
    \end{subfigure}
    \caption{For control room parameters only: (a) Target distribution for $\beta_{\theta,\mathrm{ped}}$ in the H-mode all scenario. (b) Predicted vs.\ actual on the test set ($R^2=0.674$). (c) SHAP analysis.}
    \label{fig:pedestalfits_betaped_controlroom}
\end{figure*}

\begin{figure*}[!tb]
    \centering    
    \begin{subfigure}[t]{0.36\textwidth}
    \centering
    \includegraphics[width=\textwidth]{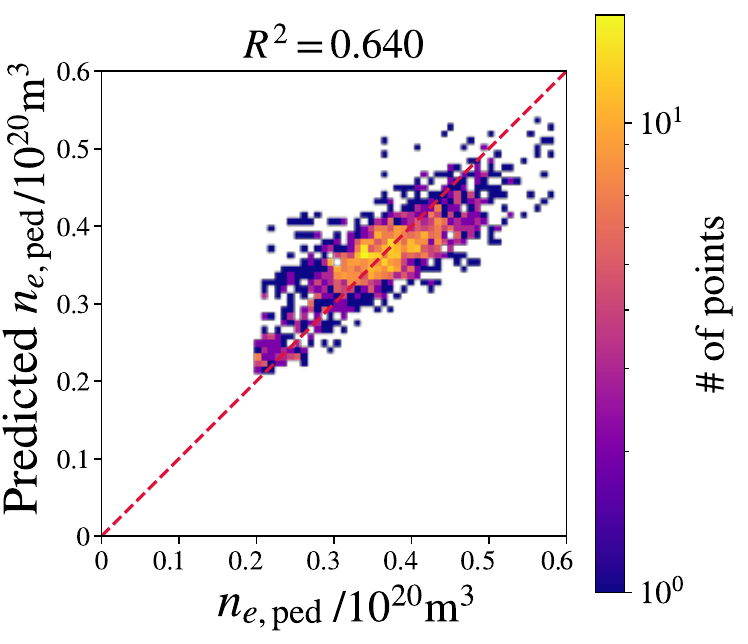}
    \caption{}
    \end{subfigure}
    \begin{subfigure}[t]{0.36\textwidth}
    \centering
    \includegraphics[width=\textwidth]{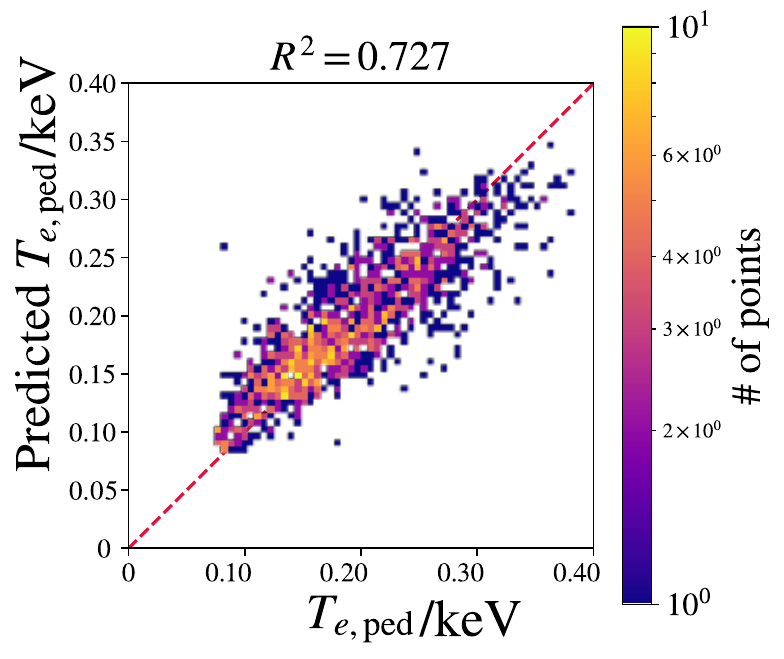}
    \caption{}
    \end{subfigure}
    \begin{subfigure}[t]{0.36\textwidth}
    \centering
    \includegraphics[width=\textwidth]{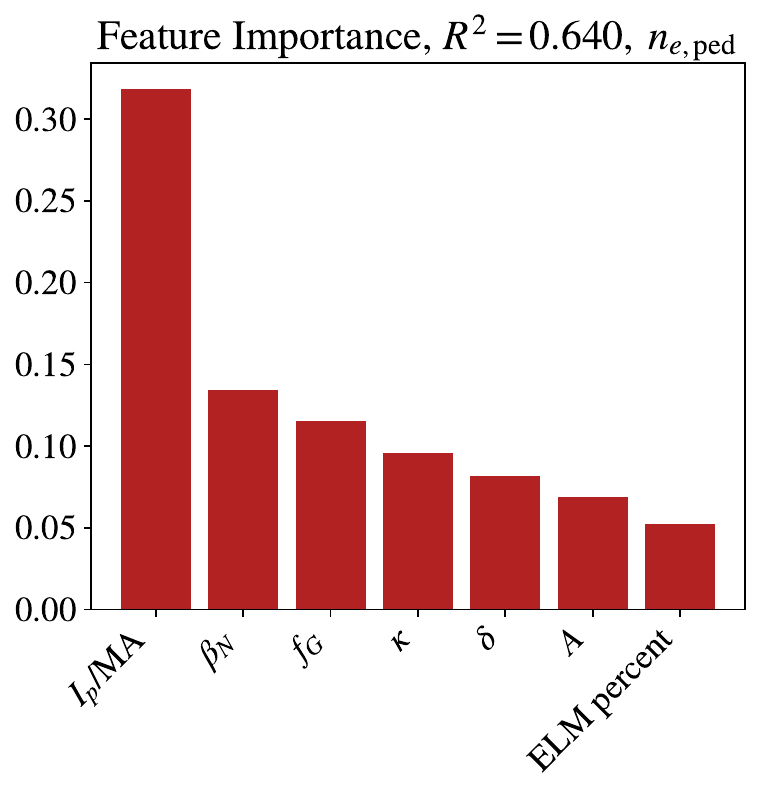}
    \caption{}
    \end{subfigure}
    \begin{subfigure}[t]{0.36\textwidth}
    \centering
    \includegraphics[width=\textwidth]{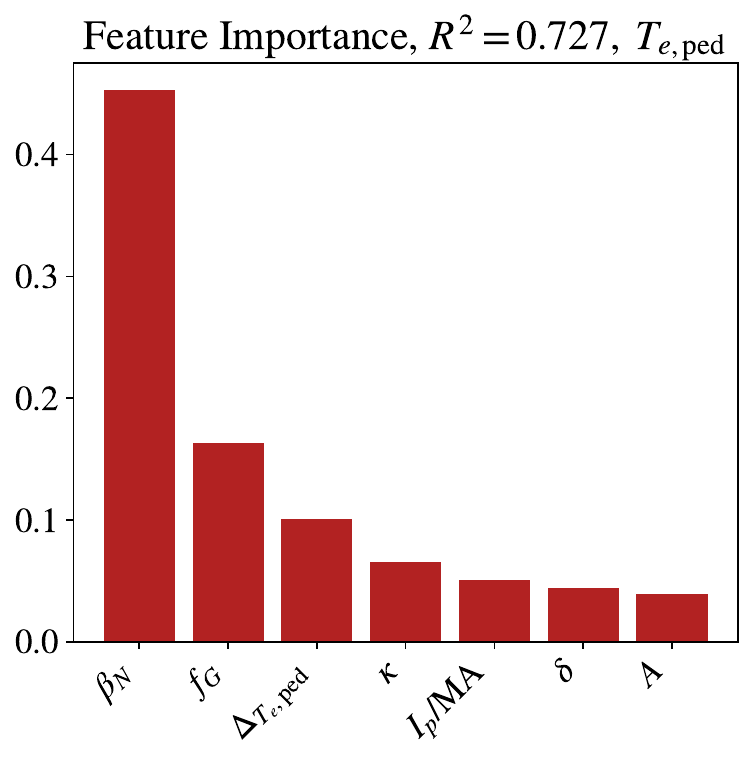}
    \caption{}
    \end{subfigure}
    \caption{Random Forest predictions for (left) pedestal density $n_{e,\mathrm{ped}}$ and (right) pedestal temperature $T_{e,\mathrm{ped}}$ in H-mode all. Top row: target distributions; bottom row: predicted vs.\ actual.}
    \label{fig:pedestalfits_ne_ped_Te_ped}
\end{figure*}

We first train an RF model to predict $\beta_{\theta,\mathrm{ped}}$ while excluding any direct pedestal-height measurements that would trivially correlate, such as $p_{e,\mathrm{ped}}$, $n_{e,\mathrm{ped}}$, or $T_{e,\mathrm{ped}}$. 

\Cref{fig:betapedtheta_RF} shows the RF predictions for $\beta_{\theta,\mathrm{ped}}$ in the three filter categories, with $R^2\approx 0.71$--0.76—significantly better than linear approaches in \Cref{fig:MASTUdatabase_full}. Feature-importance analysis reveals that $\beta_N$, $\kappa$, $\Delta_{\mathrm{ped}}$, and $f_G$ are the top predictors, underscoring how both global and pedestal-specific quantities affect pedestal height. In particular, $\beta_N$ consistently dominates the prediction, with roughly five times the weight of the next most important parameter, $\kappa$. Because the causal relation between $\beta_{\theta,\mathrm{ped} }$ and $\beta_N$ might be expected, we also train an RF model for predicting $\beta_{\theta,\mathrm{ped} }$ excluding $\beta_N$ as an input parameter. This degrades the model's predictive ability however, decreasing $R^2$ from $R^2 = 0.755$ including $\beta_N$ to $R^2 = 0.585$ without $\beta_N$ -- these results are reported in \Cref{app:noBetaN}.

To understand each parameter’s effect, we employ SHAP analysis, which assigns each feature a contribution for each prediction. This analysis shown in \Cref{fig:shap_RF_Model_betaped} suggests increasing $\beta_N$, $\kappa$, pedestal widths, $\delta$, and ELM percent raises $\beta_{\theta,\mathrm{ped}}$, whereas increasing $f_G$, $I_p$, and $A$ slightly reduces $\beta_{\theta,\mathrm{ped}}$, though these latter influences are smaller. Some of these dependencies were, and some were not obvious already from \Cref{fig:MASTU_database_histograms}. We also note that aspect ratio $A$ covered a small range in the database, and $I_p$ had three main value ranges, with $I_p =$0.75 MA plasmas dominating.

\subsection{Control Room Parameters}
\label{subsec:controlroom}

In many devices, only engineering inputs can be selected in real time, so it is practical to predict pedestal height $\beta_{\theta,\mathrm{ped}}$ using only `control room parameters’: the $P_\mathrm{beam}$ parameters, $I_p$, $\kappa$, $\delta$, $A$, $f_G$, and time within the discharge. We keep the $f_G$ parameter to approximate fueling control. As shown in \Cref{fig:pedestalfits_betaped_controlroom}, restricting inputs to these seven variables still produces a reasonable $R^2\approx 0.674$, only slightly below the $R^2=0.755$ with the full parameter set. Feature importance in \Cref{fig:pedestalfits_betaped_controlroom}(b) shows that $I_p$, $\delta$, $f_G$, and $\kappa$ play stronger roles than $P_\mathrm{beam}$ or A. The SHAP results in \Cref{fig:pedestalfits_betaped_controlroom}(c) indicate that high $I_p$ can drive $\beta_{\theta,\mathrm{ped}}$ down, possibly reflecting its inverse relation with $\beta_{N}$ or correlations with other unmodeled factors. Intermediate $f_G$ values tend to enhance $\beta_{\theta,\mathrm{ped}}$, whereas very small or large $f_G$ reduce it. Late-discharge times also tended to exhibit lower pedestal height in MAST-U, while stronger shaping ($\delta$ and $\kappa$) raises it.

\subsection{\texorpdfstring{$n_{e,\mathrm{ped}}$}{Lg} and \texorpdfstring{$T_{e,\mathrm{ped}}$}{Lg} Prediction} \label{subsec:nepedTepedprediction}

We also build separate RF models for the density pedestal $n_{e,\mathrm{ped}}$ and temperature pedestal $T_{e,\mathrm{ped}}$ heights separately, excluding direct inputs of the other to avoid trivial correlations. \Cref{fig:pedestalfits_ne_ped_Te_ped} shows that $R^2=0.64$ for $n_{e,\mathrm{ped}}$ and $R^2=0.73$ for $T_{e,\mathrm{ped}}$. Feature-importance analysis indicates $n_{e,\mathrm{ped}}$ depends strongly on $I_p$, $\beta_N$, and $f_G$, and $\kappa$, whereas $T_{e,\mathrm{ped}}$ is much less strongly influenced by $I_p$, but still strongly influenced by $\beta_N$ and $f_G$, and also $\Delta_{T_e,\mathrm{ped}}$.

We also include SHAP diagrams for $n_{e,\mathrm{ped}}$ and $T_{e,\mathrm{ped}}$ in \Cref{fig:pedestalfits_SHAP_diagrams}, shown in \Cref{app:nepedTepedSHAP} for brevity. Consistent with \Cref{fig:pedestalfits_ne_ped_Te_ped}, $n_{e,\mathrm{ped}}$ and $T_{e,\mathrm{ped}}$ have different parameter dependencies, perhaps most notably that $n_{e,\mathrm{ped}}$ only depends very weakly on pedestal width \cite{Clark2025} -- this observation might partly explain findings in an earlier section showing that $\beta_{\theta,\mathrm{ped} }$ is not well-characterized by $\Delta_\mathrm{ped}$ alone. These findings reinforce that spherical tokamak pedestals depend on multiple coupled parameters. While not visualized in this work, we found that limiting the inputs for $n_{e,\mathrm{ped}}$ and $T_{e,\mathrm{ped}}$ models to just control room inputs yields comparable values of $R^2$.

\begin{figure}[!bt]
    \centering
    \begin{subfigure}[t]{0.50\textwidth}
    \includegraphics[width=\textwidth]{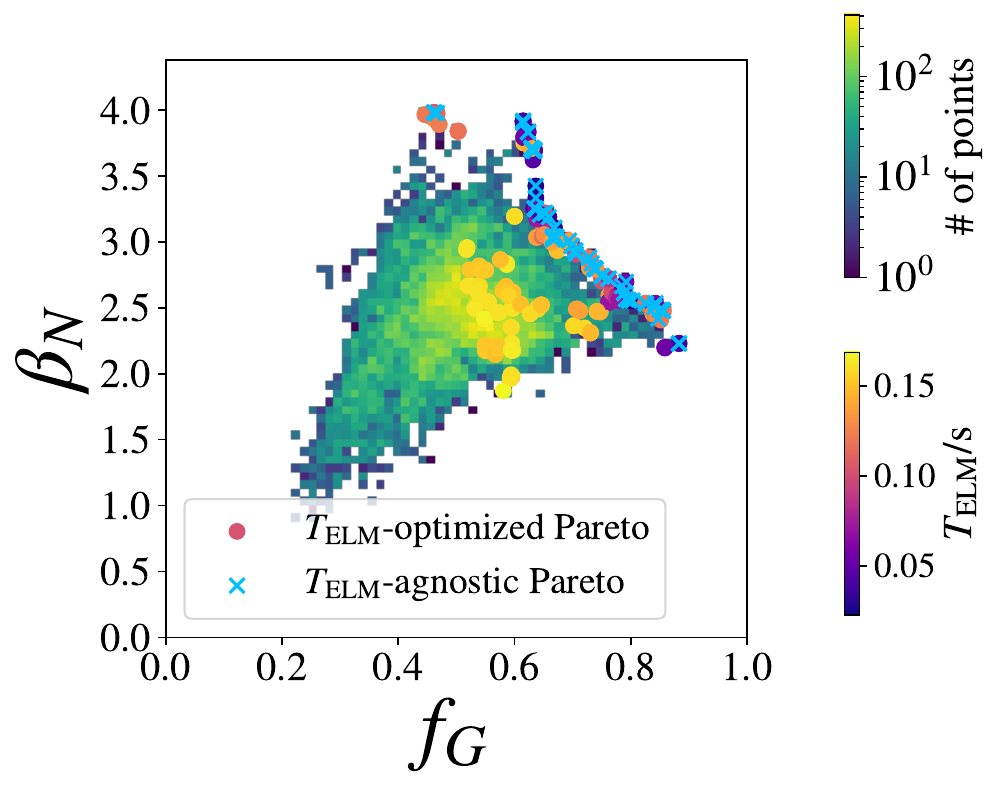}
    \end{subfigure}
    \caption{Pareto front in $(\beta_N,\,f_G)$ space, either including the objective time-to-next-ELM ($T_\mathrm{ELM}$) in `$T_\mathrm{ELM}$-optimized` or excluding it in `$T_\mathrm{ELM}$-agnostic`.}
    \label{fig:multi_param_opt}
\end{figure}

\section{Multi-Objective Pareto Optimization}
\label{sec:Pareto_optimization}

At this point, we shift focus from predicting pedestal properties to learning how to optimize plasma performance using HIPED. The goal of this section is to provide a systematic answer to the question:``Given a database of experimental data from tokamak H-mode discharges, how does one find the `best' discharges, how does one find the parameters corresponding to the `best' discharges, and how would one achieve even better performance?''

A key advantage of database analysis is multi-objective optimization. Experiments must often balance multiple goals (e.g. maximizing performance while mitigating edge fluxes). Pareto optimization (PO) identifies parameter sets for which no single objective can be further improved without worsening at least one other. By constructing Pareto fronts, we map out tradeoffs among competing objectives in spherical tokamak experiments. According to the priorities of a given experiment, the weightings of the relative objectives can be adjusted. This provides a quantitative approach for prioritizing different objectives.

In this section, we find the Pareto front using synthetic data based on the underlying database -- Pareto fronts are found for four Cases with different optimization criteria. We then calculate how to further improve high-performance discharges. While we do not explicitly optimize for pedestal width and height, we do optimize for the time until the next ELM. Therefore, pedestal physics is included in our Pareto optimization through the lens of minimizing ELM frequency.

\subsection{Pareto Fronts}

We calculate the Pareto front with four objectives: maximizing $\beta_N$, $f_G$, the line-averaged density $\langle n_e\rangle_L$, and the time until the next ELM $T_{\mathrm{ELM}}$. Note that $\langle n_e\rangle_L$ can be obtained from $f_G$, $I_p$, and the plasma minor radius \cite{Greenwald1988} -- since the minor radius varies little between discharges, by using $\langle n_e\rangle_L$ we are not adding a new independent input variable but rather using a different form of $I_p$. By sampling the broader input-parameter space (50,000 synthetic samples based on the underlying MAST-U database), we compute the predicted values of each objective from our RF models and identify Pareto-optimal points in 4D. We also consider points that do not optimize at all for $T_{\mathrm{ELM}}$, which is more appropriate for ELMy H-modes. Solutions that exclude $T_{\mathrm{ELM}}$ can achieve higher $\beta_N$, $f_G$, or $\langle n_e\rangle_L$ but typically exhibit shorter ELM-free intervals, and vice versa.

The Pareto optimization results are shown in \Cref{fig:multi_param_opt}. We plot the number of points in $\beta_N$-$f_G$ space for the entire Pareto parameter grid using the blue/green/yellow colormap. These points are reflective of the properties of discharges we retained in the MAST-U database. 

On top of the number of points, circular markers with varied colors are Pareto-front points for the `$T_\mathrm{ELM}$-optimized' case, which is where we used all four of the variables $\beta_N$, $f_G$, $\langle n_e \rangle_L$, and $T_{\mathrm{ELM}}$ in calculating the Pareto front. Each point corresponds to a Pareto optimum -- a point where no single objective (in this work, any one of $\beta_N$, $f_G$, $\langle n_e \rangle_L$, and $T_{\mathrm{ELM}}$) can be further improved without worsening at least one other. Collectively, we refer to these points as the Pareto front. However, more information is required to select which point specific on the Pareto front is optimal -- this extra information is an importance weighting for each of the four optimization objectives. We will implement this extra information in the next section. Up until now, \Cref{fig:multi_param_opt} only shows the space of all Pareto-optimal points. It is up to the user to prioritize which of the four objectives are most important.

The marker color in \Cref{fig:multi_param_opt} shows the time until the next ELM. Notably, markers with higher $T_{\mathrm{ELM}}$ values have smaller $f_G$ and $\beta_N$ values, suggesting that there is a trade-off between the time until the next ELM and high $f_G$ and $\beta_N$ values. The blue cross markers for the `$T_\mathrm{ELM}$-agnostic' case only used  $\beta_N$, $f_G$, and $\langle n_e \rangle_L$ in calculating the Pareto front, excluding $T_{\mathrm{ELM}}$. As expected, the `$T_\mathrm{ELM}$-agnostic' markers lie on top of the `$T_\mathrm{ELM}$-optimized' markers with the lowest times until the next ELM. This is because the `$T_\mathrm{ELM}$-agnostic' Pareto front does not consider $T_{\mathrm{ELM}}$ as a quantity to maximize -- the additional optimization dimension of $\langle n_e \rangle_L$ is not shown, but we will consider it in the next section.

\subsection{MAST-U Pareto-Optimal Shots}
\label{sec:pareto}

We next find which actual MAST-U shots and time points lie closest to the Pareto fronts through a Multi-Criteria Decision-Making (MCDM) approach \cite{Taherdoost_2023}. Each objective (e.g.\ $\beta_N$, $f_G$, $T_\mathrm{ELM}$, $\langle n_e\rangle_L$) is assigned a weight, and a single scalar score is formed. The point with the highest score is deemed the best match given the specified weighting. By sampling a large space of input parameters, we identify a set of Pareto-optimal solutions. Solutions that ignore $T_{\mathrm{ELM}}$ will achieve higher $\beta_N$, $f_G$, and $\langle n_e \rangle_L$, but at the expense of shorter ELM-free intervals. In contrast, including $T_{\mathrm{ELM}}$ yields different operating points that prolong the time to ELM onset, which is essential for achieving ELM-free regimes. The optimization is performed for the H-mode all filter, but without the ELM-cycle filtering. This increases the number of individual timeslices to 9374.

\begin{figure}[!bt]
    \centering
    \begin{subfigure}[t]{0.50\textwidth}
    \includegraphics[width=\textwidth]{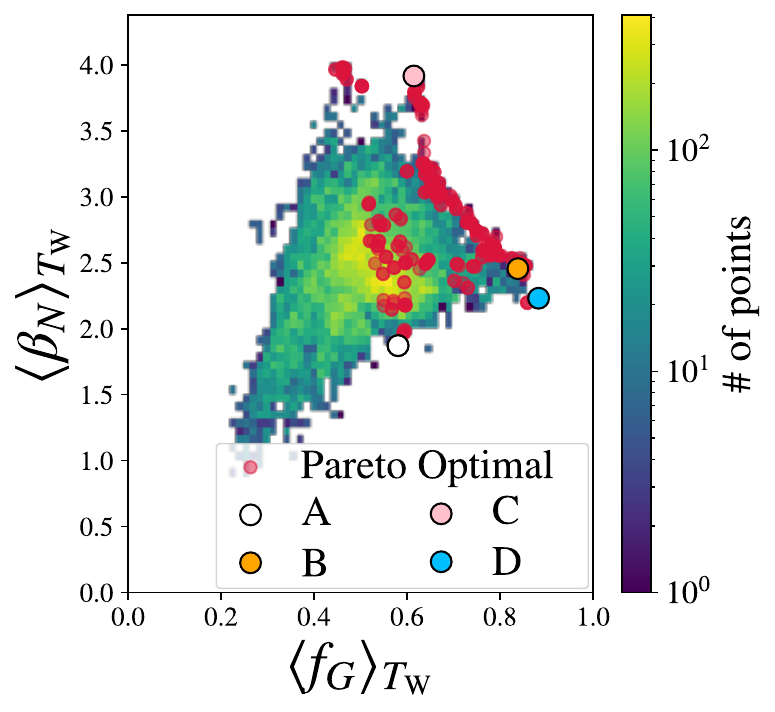}
    \end{subfigure}
    \begin{subfigure}[t]{0.50\textwidth}
    \includegraphics[width=\textwidth]{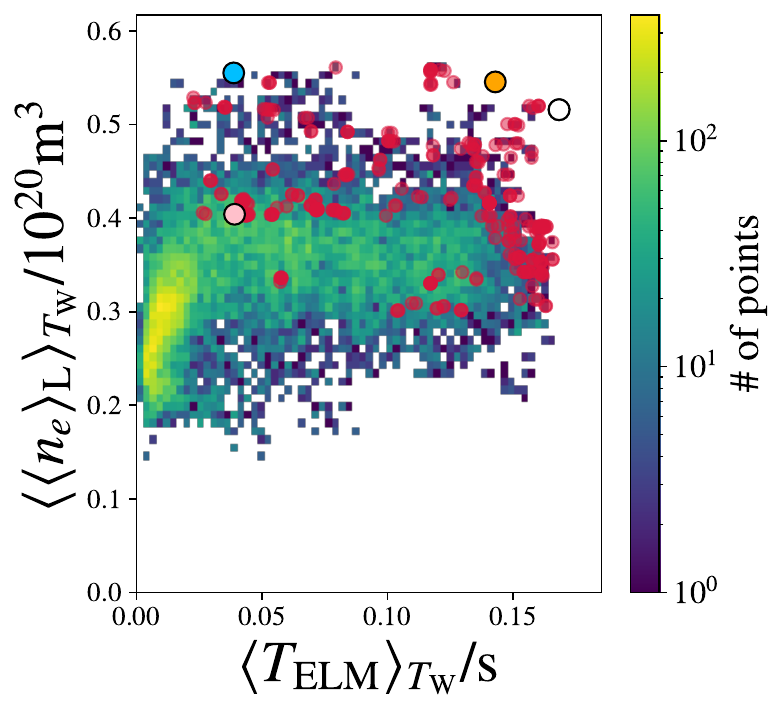}
    \end{subfigure}
    \caption{Example Pareto front in (a)~$(\beta_N,f_G)$ and (b)~$(T_\mathrm{ELM},\langle n_e\rangle_L)$, with various weightings for MCDM. Different points become optimal under different objective priorities.}
    \label{fig:pareto_further}
\end{figure}

\begin{figure*}[bt!]
    \centering
    \begin{subfigure}[t]{0.48\textwidth}
    \includegraphics[width=\textwidth]{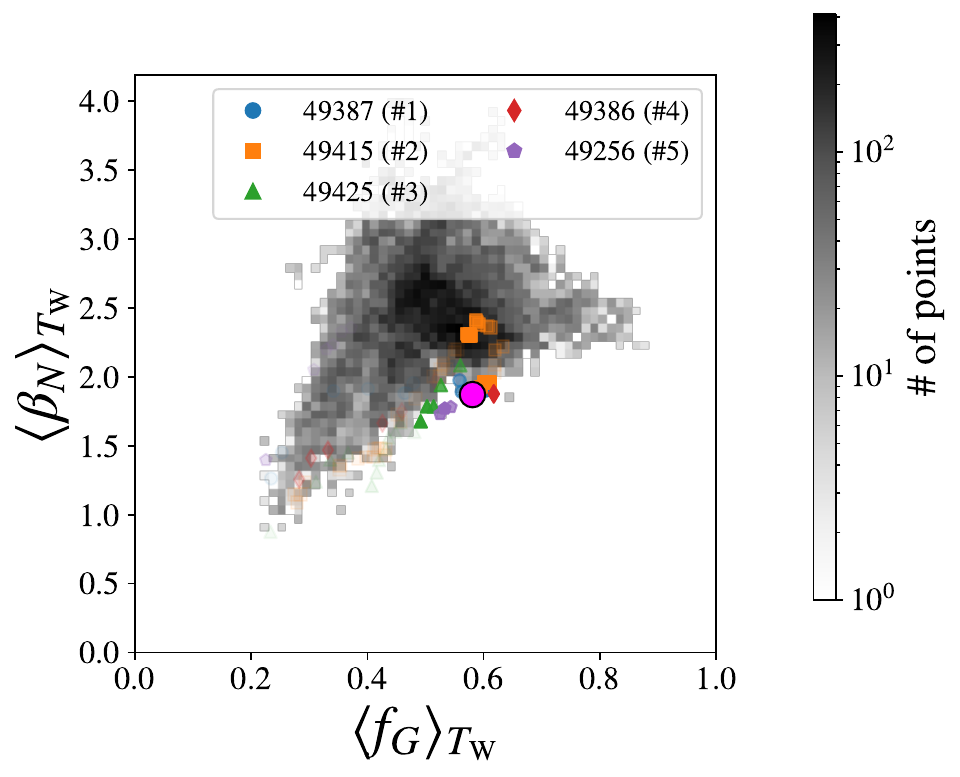}
    \caption{Case A.}
    \end{subfigure}
    \begin{subfigure}[t]{0.48\textwidth}
    \includegraphics[width=\textwidth]{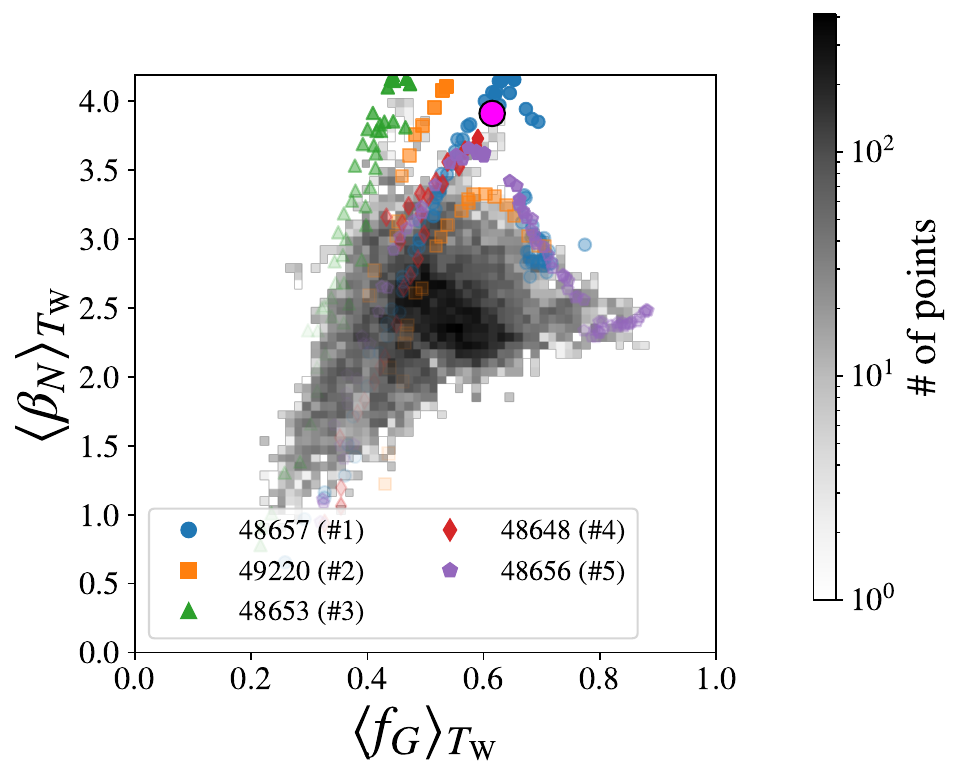}
    \caption{Case C.}
    \end{subfigure}
    \centering
    \begin{subfigure}[t]{0.48\textwidth}
    \includegraphics[width=\textwidth]{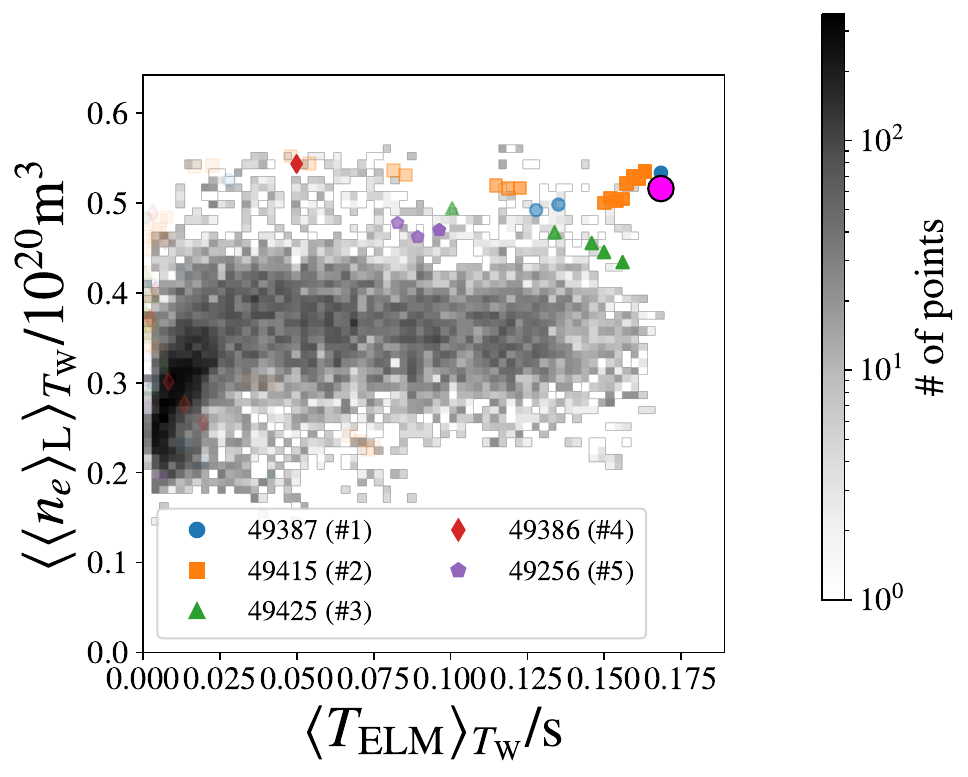}
    \caption{Case A.}
    \end{subfigure}
    \begin{subfigure}[t]{0.48\textwidth}
    \includegraphics[width=\textwidth]{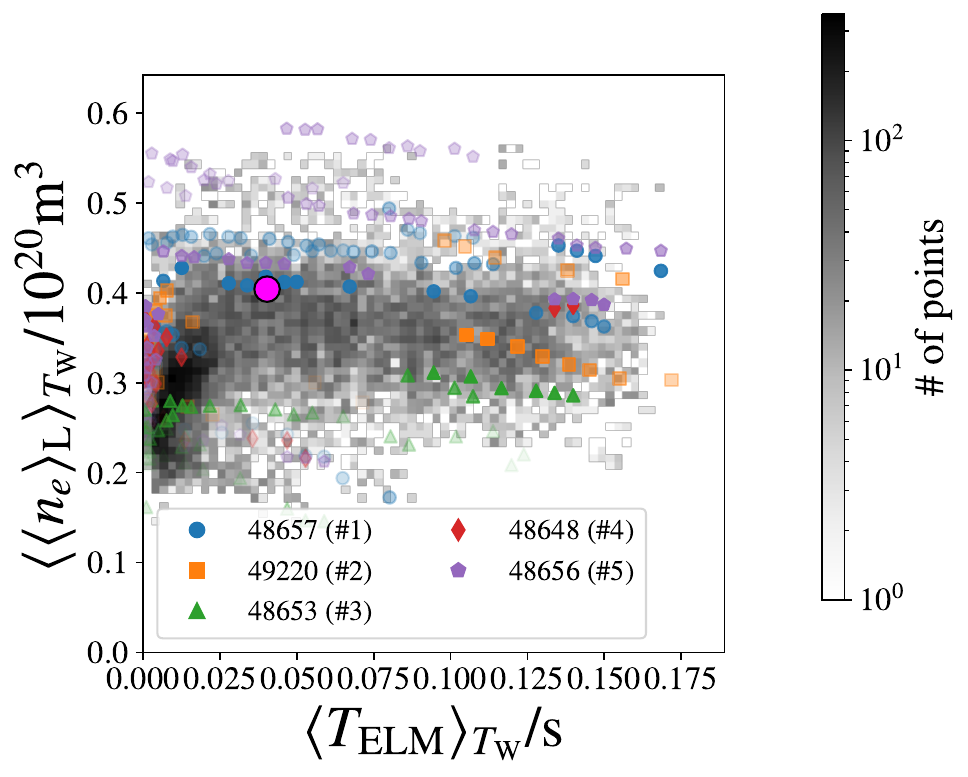}
    \caption{Case C.}
    \end{subfigure}
    \caption{Pareto-optimal MAST-U points for Cases A and C. The marker transparency is inversely proportional to the square of the MCDM score normalized between 0 and 1. A higher transparency reflects a lower MCDM score. Purple marker indicates MCDM point for the Case.}
    \label{fig:pareto_further2}
\end{figure*}

To identify specific MAST-U shots whose timeslices lie closest to the Pareto front, we employ MCDM. In this approach, the MAST-U discharge with the highest score $S_j$
\begin{equation}
S_j = \sum_{i=1}^4 w_i  \hat{s}_{ij}.
\label{eq:MCDMscore}
\end{equation}
is selected as the `optimal' choice. Here, $j$ is an index over all the equilibria in our H-mode all scenario, $i$ is an index over the features that we are optimizing, $w_i$ is a weight, and $\hat{s}_{ij}$ is a normalized score. In this work, we will consider four features corresponding to the variables used for PO in the previous subsection: \( i = 1 \leftrightarrow  f_G \), \( i = 2 \leftrightarrow \beta_N \),  \( i=3 \leftrightarrow T_\mathrm{ELM} \), and \( i=4 \leftrightarrow \langle n_e \rangle_L \). The weights satisfy $\sum_{i=1}^4 w_i = 1$.
In our study, these weights are denoted as \( w_\mathrm{f_G} \), \( w_{\beta_N} \), \( w_{T_\mathrm{ELM}} \), and \( w_{\langle \mathrm{n_e} \rangle_L} \). The normalized score is given by
\begin{equation}
\hat{s}_{ij} = \frac{s_{ij}}{\max_j(s_{ij})},
\label{eq:normalizedscore1}
\end{equation}
where the maximum $\max_j(s_{ij})$ is taken over all values in the database for each input value. The unnormalized score $s_{ij}$ is the value of each input parameter, i.e. for shot $j$, $ s_{1j} = f_G $, $ s_{2j} = \beta_N $, \( s_{3j} = T_\mathrm{ELM} \), and \( s_{4j} = \langle n_e \rangle_L \). The discharge and corresponding timeslice with the highest overall score \( S_j \) is the discharge whose timeslice is closest to the Pareto front given the specified weights.

To favor more sustained (rather than transient) performance, we also apply a time-window averaging approach. We replace each objective $s_{ij}$ with its average over a time window $T_\mathrm{window}$ centered on the candidate time point, thus selecting discharges that maintain high performance for longer intervals. We denote the time-window average for a quantity $g$ centered at time $t_0$ as
\begin{equation}
\langle g \rangle_{T_\mathrm{window}} \equiv \frac{ \int_{t_0 - T_\mathrm{window}/2}^{t_0 + T_\mathrm{window}/2} g \; dt }{T_\mathrm{window}}.
\end{equation}
Therefore, the normalized score in the MCDM approach in \Cref{eq:normalizedscore1} becomes
\begin{equation}
\hat{s}_{ij} = \frac{ \langle s_{ij} \rangle_{T_\mathrm{window}} }{ \max_j\left(\langle s_{ij} \rangle_{T_\mathrm{window}} \right)}.
\label{eq:normalizedscore2}
\end{equation}
When finding the discharge and timeslice with the highest score $S_j$, we use the time-averaged normalized score in \Cref{eq:normalizedscore2}. This ensures that the high performance attributed to a discharge at the Pareto front is not a transitory state but is more long-lived. In this work we use $T_\mathrm{window}= 0.05$s.

We explore four weight combinations (Cases A, B, C, D) as summarized in \Cref{table:MASTU_MCDMweightings}.  Case A heavily prioritizes long time-to-ELM, Case B prioritizes all quantities equally, Case C strongly prioritizes high $f_G$, and Case D strongly prioritizes high $\beta_N$. \Cref{fig:pareto_further} shows how the resulting optimal points evolve in $(\beta_N,f_G)$ and $(T_\mathrm{ELM},\langle n_e\rangle_L)$ spaces under different weights. Shots that prioritize delaying ELMs typically sacrifice $\beta_N$ or $f_G$, but sustain higher $T_\mathrm{ELM}$. For simplicity, in the main text we focus on comparing Cases A and C. We show results for all four cases: A,B,C, and D in \Cref{app:pareto_optimal}.

\begin{table}[bt!]
\centering
\begin{tabular}{|c|c|c|c|c|}
\hline
\cline{2-5}
\textbf{Case} & $w_{T_\mathrm{ELM}}$ & $w_{f_G}$ & $w_{\beta_N}$ & $w_{\langle n_e\rangle_L}$ \\
\hline
A & 0.70 & 0.10 & 0.10 & 0.10 \\
\hline
B & 0.25 & 0.25 & 0.25 & 0.25 \\
\hline
C & 0.00 & 0.10 & 0.80 & 0.10 \\
\hline
D & 0.00 & 0.80 & 0.10 & 0.10 \\
\hline
\end{tabular}
\caption{Weighting parameters $w_{T_\mathrm{ELM}}$, $w_{f_G}$, $w_{\beta_N}$, and $w_{\langle n_e\rangle_L}$ for the four Cases considered in this work. Weighting parameters place different emphasis on the optimization parameters $T_\mathrm{ELM}$, $f_G$, $\beta_N$, and $\langle n_e\rangle_L$.}
\label{table:MASTU_MCDMweightings}
\end{table}

\begin{figure*}[bt!]
    \centering
    \begin{subfigure}[t]{0.48\textwidth}
    \includegraphics[width=\textwidth]{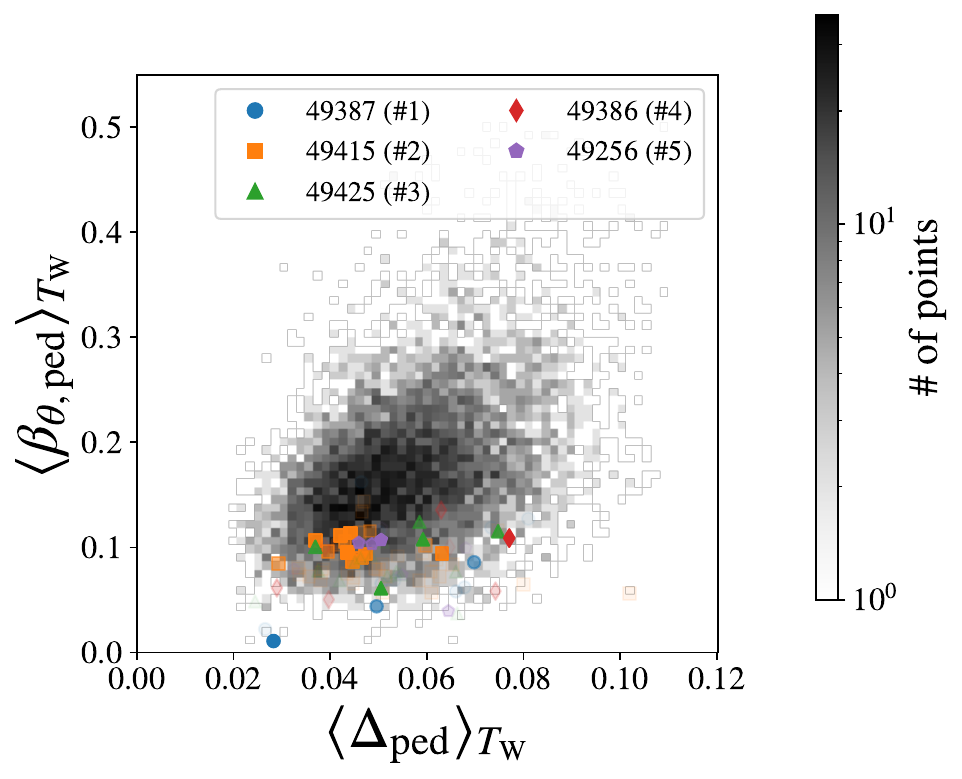}
    \caption{Case A.}
    \end{subfigure}
    \centering
    \begin{subfigure}[t]{0.48\textwidth}
    \includegraphics[width=\textwidth]{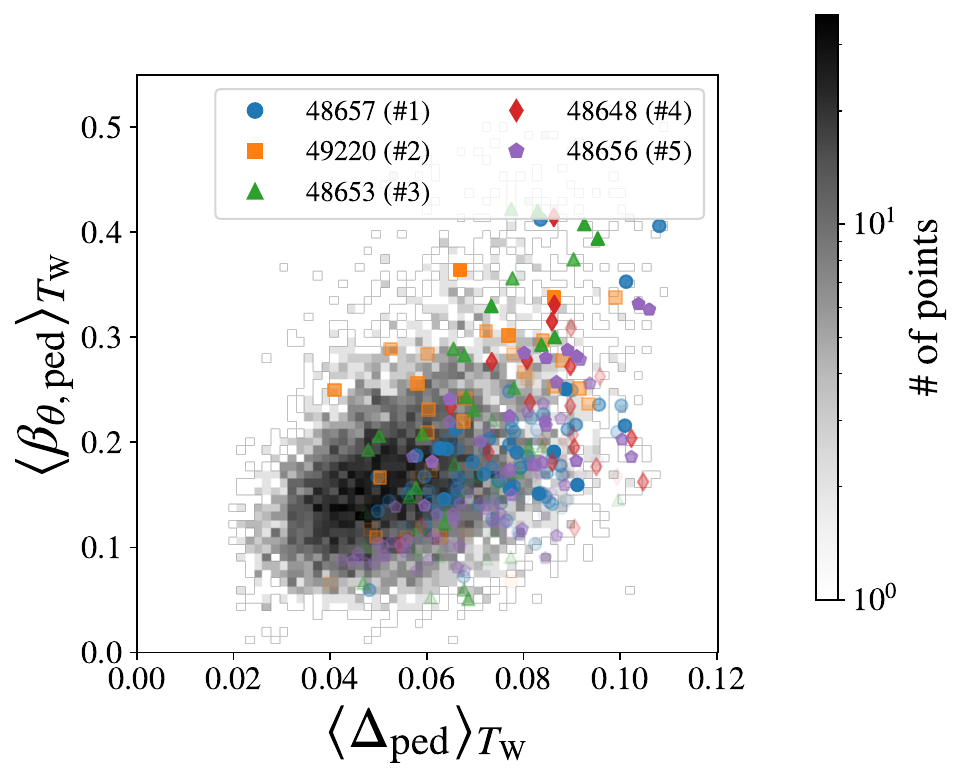}
    \caption{Case C.}
    \end{subfigure}
    \centering
    \begin{subfigure}[t]{0.48\textwidth}
    \includegraphics[width=\textwidth]{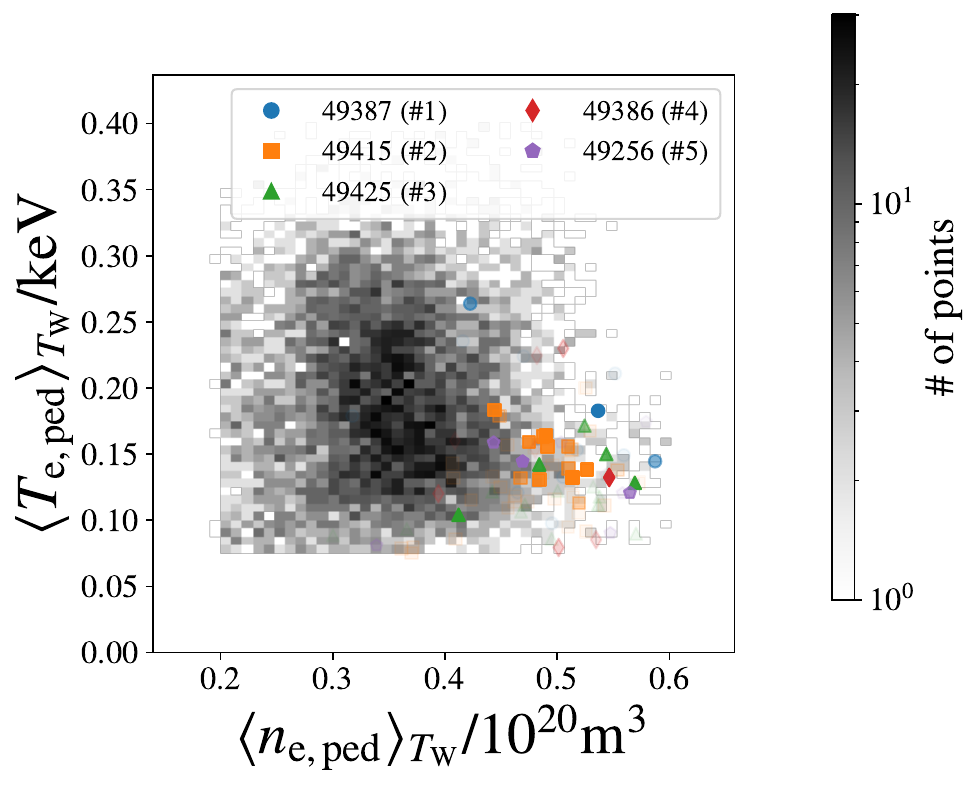}
    \caption{Case A.}
    \end{subfigure}
    \centering
    \begin{subfigure}[t]{0.48\textwidth}
    \includegraphics[width=\textwidth]{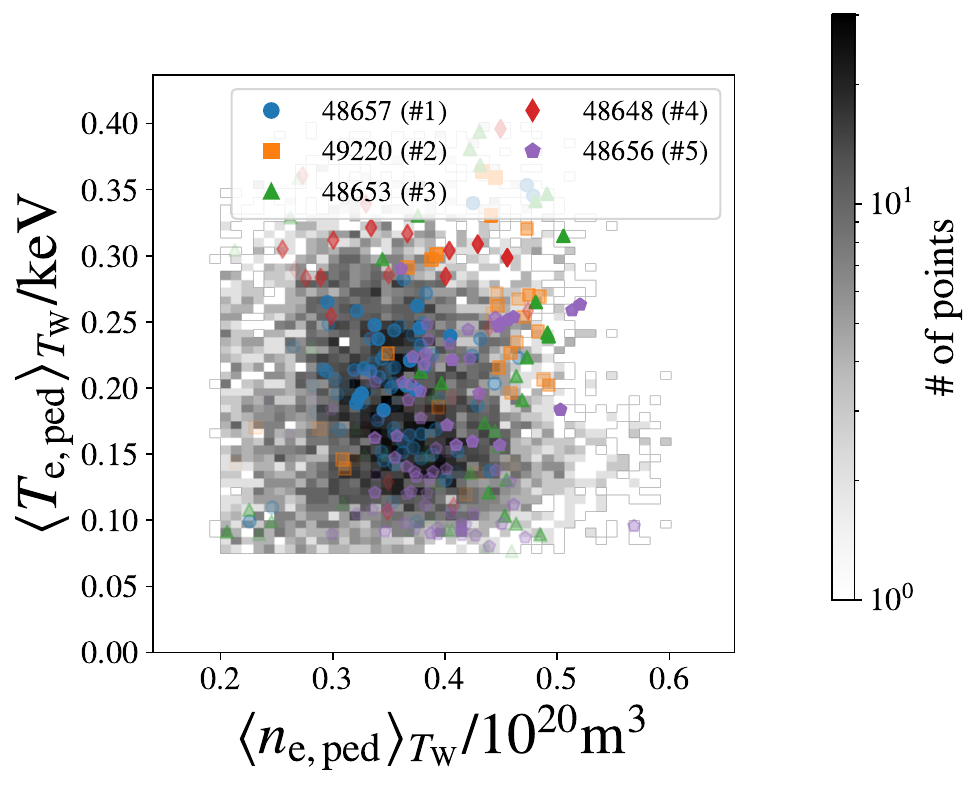}
    \caption{Case C.}
    \end{subfigure}
    \caption{Pareto-optimal MAST-U points for Cases A and C. The marker transparency is inversely proportional to the square of the MCDM score normalized between 0 and 1. A higher transparency reflects a lower MCDM score.}
    \label{fig:pareto_further3}
\end{figure*}

We plot trajectories for the five discharges closest to the Pareto front in \Cref{fig:pareto_further2} for Cases A and C. The top row of \Cref{fig:pareto_further2} shows $\beta_N$ and $f_G$: for Case A, discharges achieve lower $\beta_N$ than for Case C. However, the bottom row of \Cref{fig:pareto_further2} shows that discharges in Case A achieve simultaneously higher $T_\mathrm{ELM}$ and $\langle n_e \rangle_L$ than case C.

Next, we examine pedestal parameters for the MAST-U discharges close to each Pareto front. The pedestal parameter space trajectories differ significantly between Case A and C. Case C has noticeably higher pedestal height $\beta_{\theta,\mathrm{ped}}$, width $\Delta_\mathrm{ped}$ (first row in \Cref{fig:pareto_further3}) and temperature pedestal $T_{e\mathrm{ped}}$ (second row in \Cref{fig:pareto_further3}), whereas the density pedestal $n_{e\mathrm{ped}}$ (second row in \Cref{fig:pareto_further3}) is slightly higher for Case A.

\begin{figure*}[bt!]
    \begin{subfigure}[t]{0.43\textwidth}
    \includegraphics[width=\textwidth]{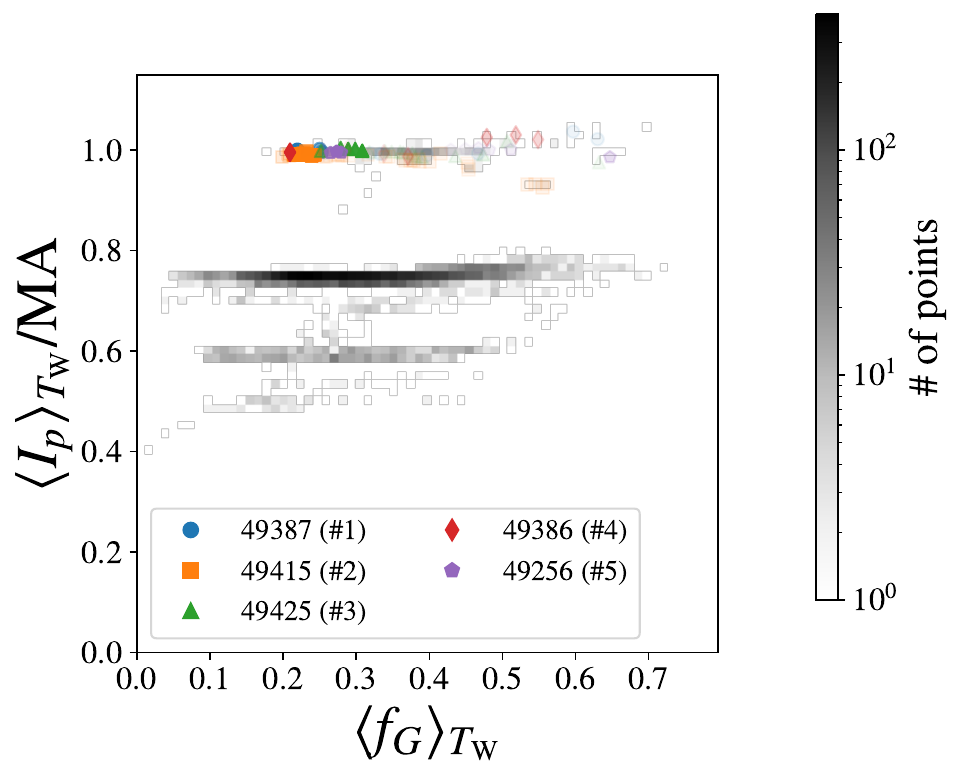}
    \caption{Case A}
    \end{subfigure}
    \begin{subfigure}[t]{0.43\textwidth}
    \includegraphics[width=\textwidth]{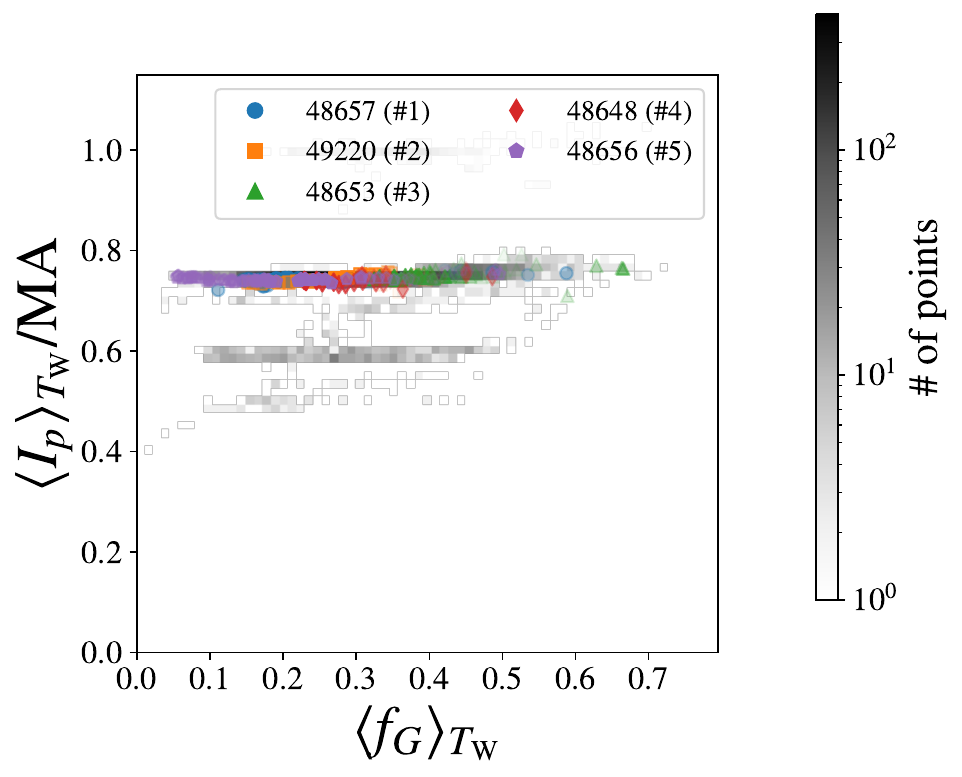}
    \caption{Case C}
    \end{subfigure}
    \begin{subfigure}[t]{0.43\textwidth}
    \includegraphics[width=\textwidth]{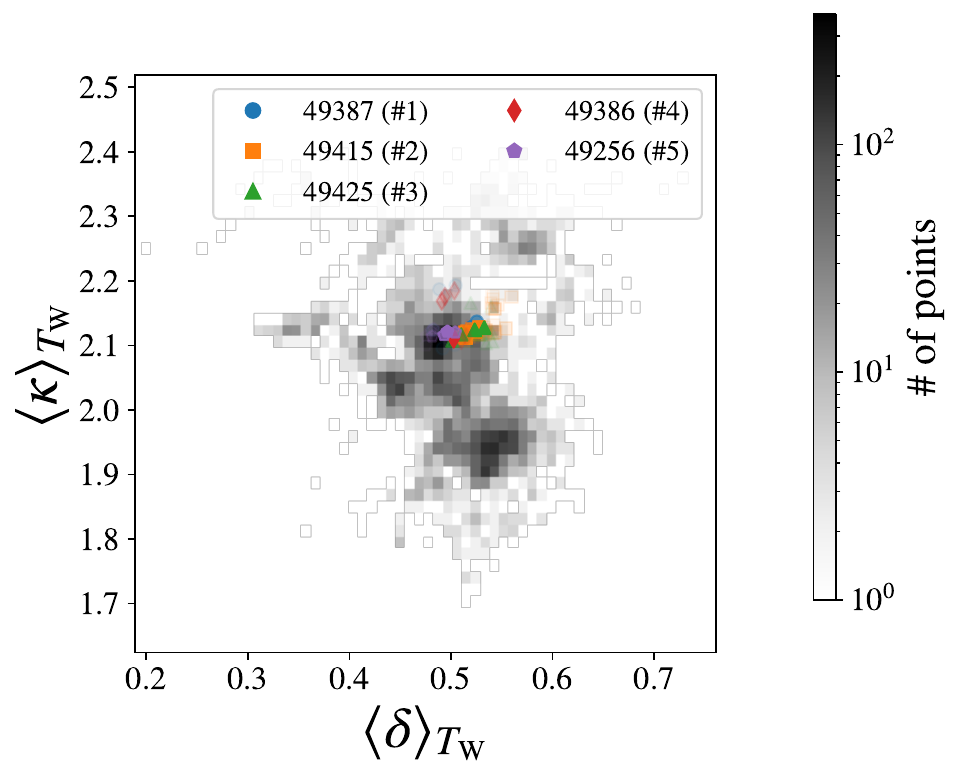}
    \caption{Case A}
    \end{subfigure}
    \begin{subfigure}[t]{0.43\textwidth}
    \includegraphics[width=\textwidth]{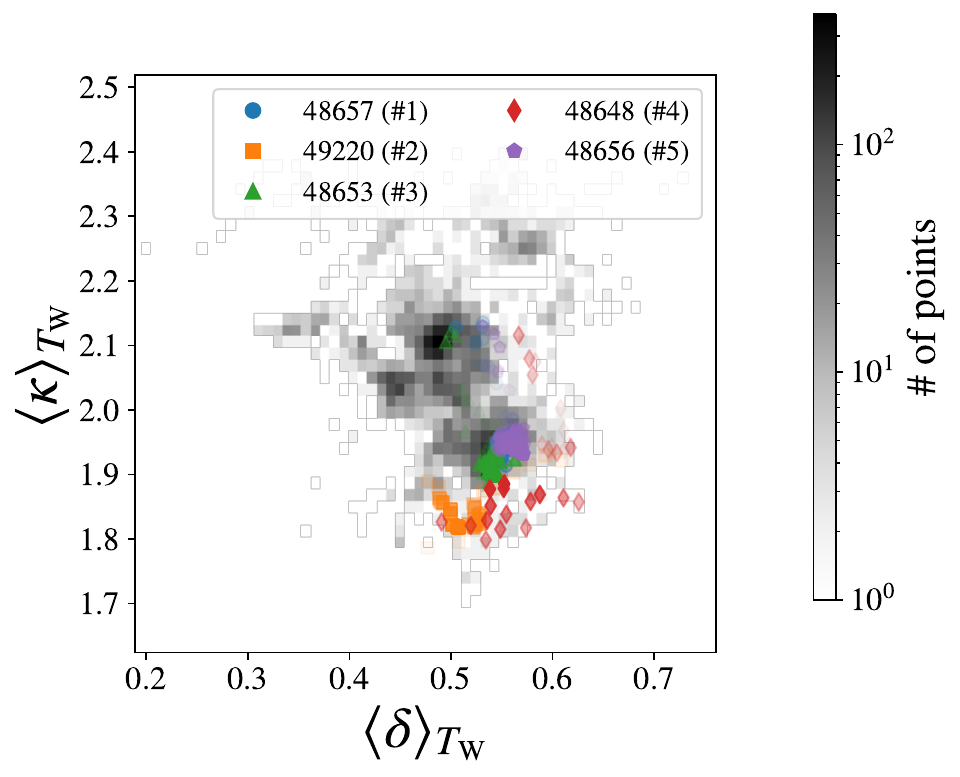}
    \caption{Case C}
    \end{subfigure}
    \begin{subfigure}[t]{0.43\textwidth}
    \includegraphics[width=\textwidth]{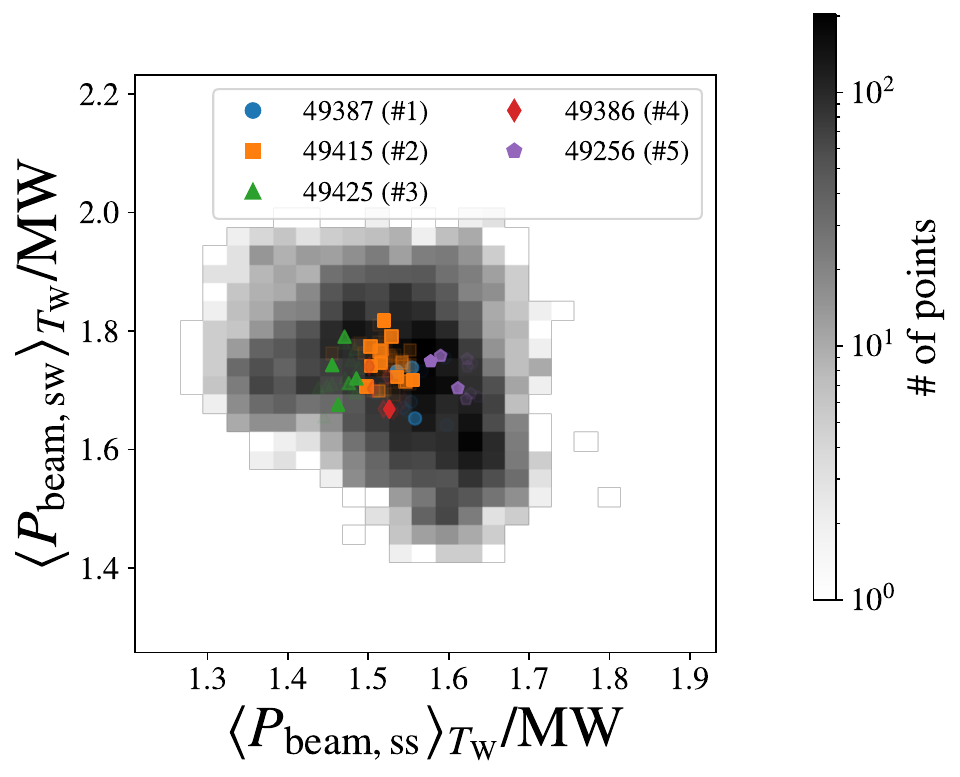}
    \caption{Case A}
    \end{subfigure}
    \begin{subfigure}[t]{0.43\textwidth}
    \includegraphics[width=\textwidth]{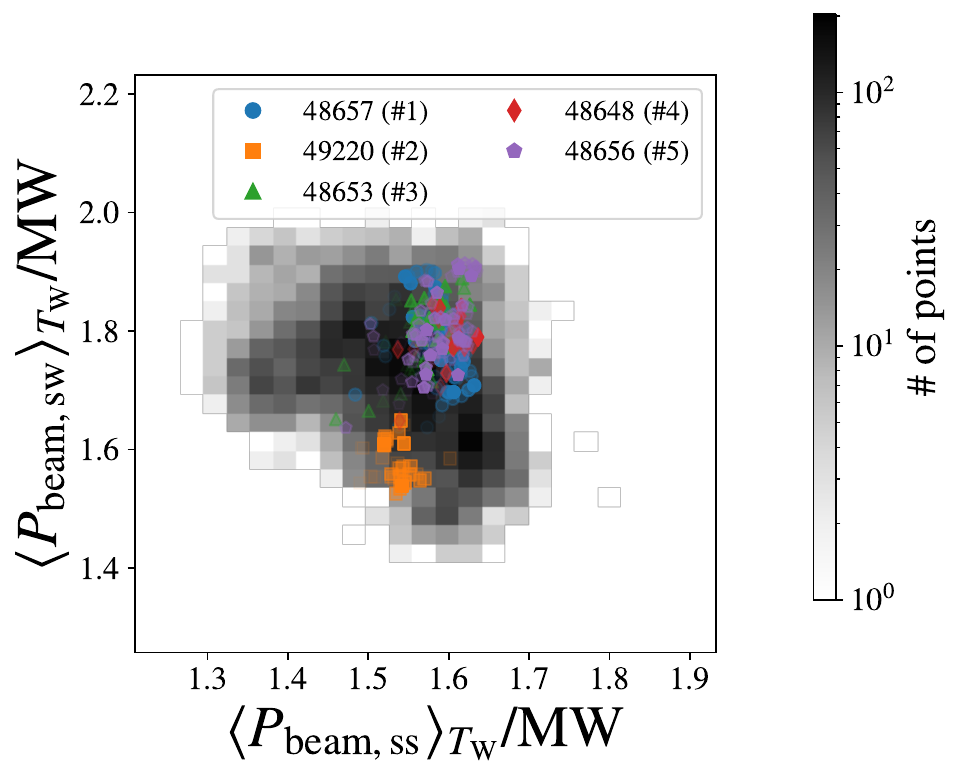}
    \caption{Case C}
    \end{subfigure}
    \caption{Control-room parameters for Pareto-optimal time points in MAST-U, Cases A and C: $(I_p,f_G)$, $(\delta,\kappa)$, $(A,P_\mathrm{beam})$, $(P_{\mathrm{beam,ss}},P_{\mathrm{beam,sw}})$. The marker transparency is inversely proportional to the square of the MCDM score normalized between 0 and 1. A higher transparency reflects a lower MCDM score.}
    \label{fig:pareto_further4}
\end{figure*}

Next, we show control room parameters for Cases A and C in \Cref{fig:pareto_further4}. The first column shows $I_p$ and $f_G$: the less ELMy discharges in Case A all have the highest $I_p$ values. The second column shows $\kappa$ and $\delta$: Case A has higher elongation and lower triangularity. The role of shaping is distinct for spherical tokamaks where ELM-free H-modes are accessible with strong shaping \cite{Parisi_2024b,Parisi_2024c,Imada2024,Imada2024b,Nelson2024c}. The result that optimizing for higher $\beta_N$ gives lower $\kappa$ is consistent with \cite{Zamkovska_2024} where it was shown that the maximum achievable $\beta_N$ in MAST-U decreased with $\kappa$ due to increased risk of disruptions. The third column shows the Southwest and South beam power, $P_\mathrm{beam,sw}$ and $P_\mathrm{beam,ss}$. Case A has lower $P_\mathrm{beam,ss}$ values. By showing the control room parameters, our approach can help experimentalists plan for future experiments.

\begin{figure*}[bt!]
    \centering
    \begin{subfigure}[t]{0.8\textwidth}
    \centering
    \includegraphics[width=1.0\textwidth]{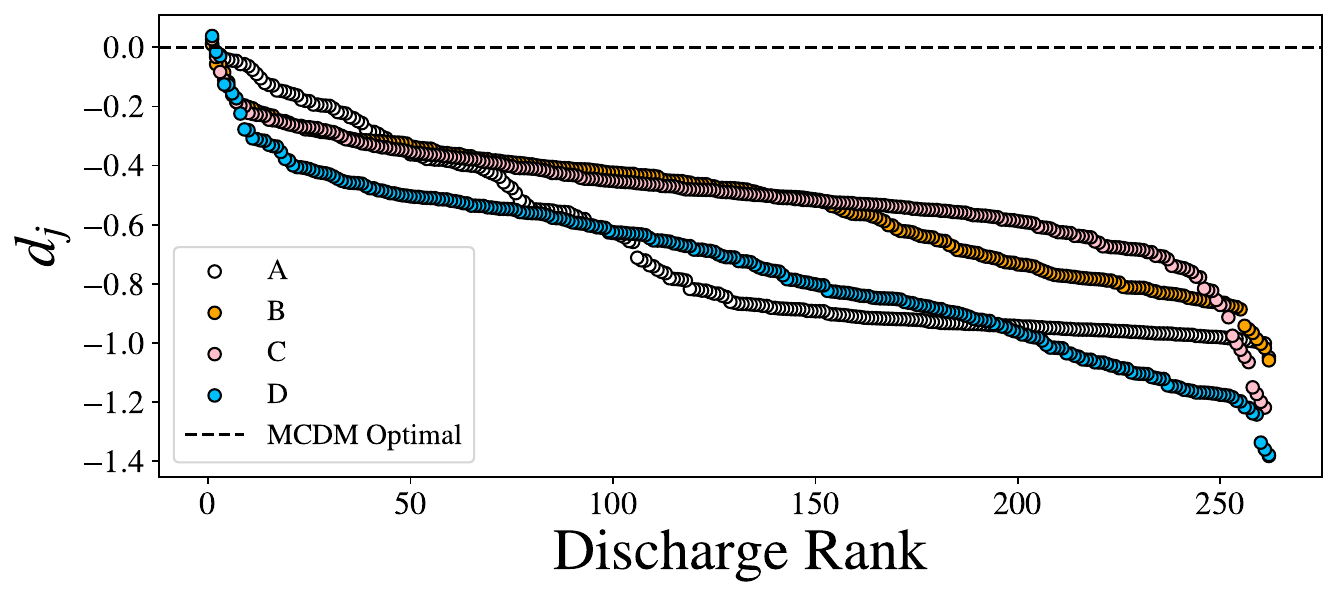}
    \end{subfigure}
    \caption{Distance from MCDM $d_j$ for MAST-U discharges for four Cases A-D. A lower value of $d_j$ indicates a poorer performance relative to the MCDM optimal for the weighting in that Case.}
    \label{fig:neighborsscore}
\end{figure*}

\subsection{Discharge Ranking and Optimization}

In this subsection, we calculate the ranking of discharges relative to optimal MCDM solutions for the four Cases. In order to measure the proximity of a discharge to the MCDM optimal point, we calculate the distance from MCDM,
\begin{equation}
    d_j = S_j - S_\mathrm{optimal},
    \label{eq:distancetoMCDM}
\end{equation}
where $S_j$ is the MCDM score defined in \Cref{eq:MCDMscore} and $S_\mathrm{optimal}$ is the score of the optimal point on the Pareto front. For each MAST-U discharge, we find the timeslice with the highest $d_j$ and report that $d_j$ value as corresponding to that particular discharge. As in the above sections, we use the time-window average for calculating $S_j$ and $S_\mathrm{optimal}$. For each of the four cases, we plot $d_j$ in \Cref{fig:neighborsscore}. A score $d_j > 0$ indicates a MAST-U discharge with a higher MCDM score than the MCDM optimal from our synthetic Pareto database, and $d_j$ indicates that the discharge is lower performance that the MCDM optimal. Curiously, while all of the Cases have one MAST-U discharge that is slightly above the Pareto front, indicated by $d_j > 0$, the falloff in high-performance discharges (for example, $d_j \gtrsim -0.2$) is very fast for Cases B,C, and D and slowest for Case A. The relatively slow falloff for case A indicates that achieving ELM-free H-modes was more common, as long as one is willing to sacrifice core plasma performance, here measured by $f_G$, $\beta_N$, and $T_\mathrm{ELM}$. The relatively fast falloff for Cases B, C, and D indicates that achieving a high MCDM score is more challenging for these weighting parameters (see \Cref{table:MASTU_MCDMweightings}) that more heavily emphasize improved core plasma performance. %

\begin{figure*}[bt!]
    \centering
    \begin{subfigure}[t]{0.68\textwidth}
    \centering
    \includegraphics[width=1.0\textwidth]{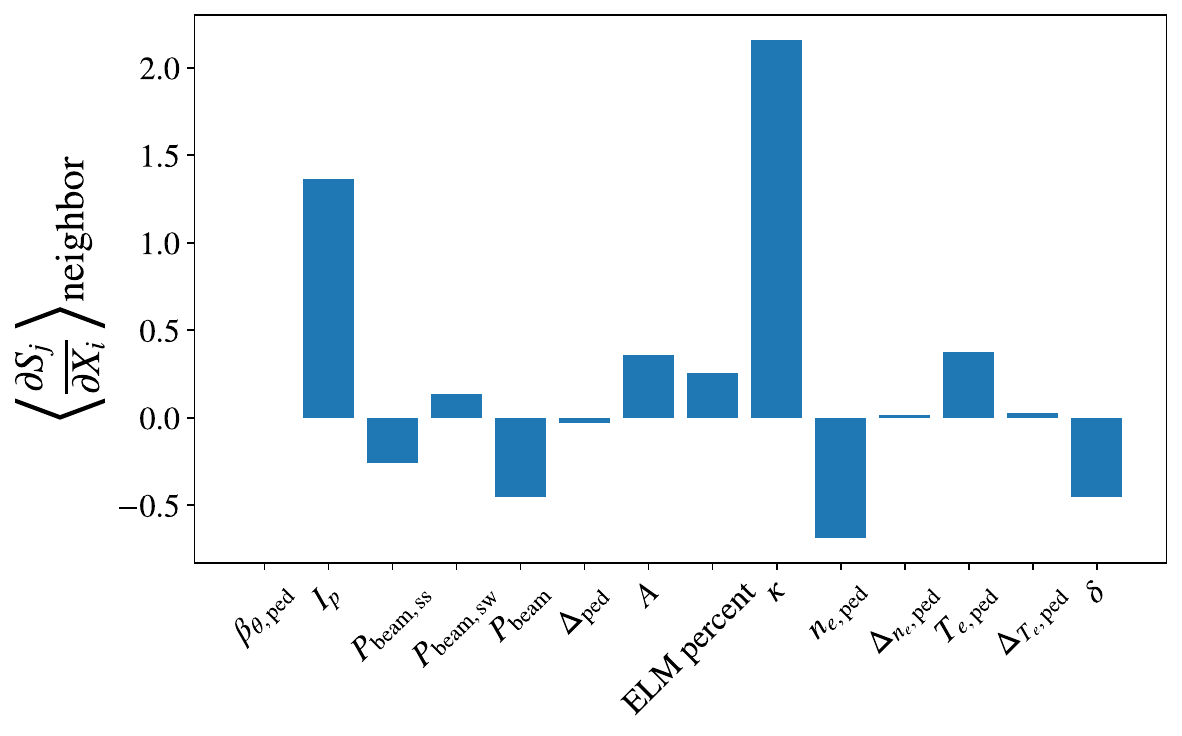}
    \caption{Case A}
    \end{subfigure}
    \begin{subfigure}[t]{0.68\textwidth}
    \centering
    \includegraphics[width=1.0\textwidth]{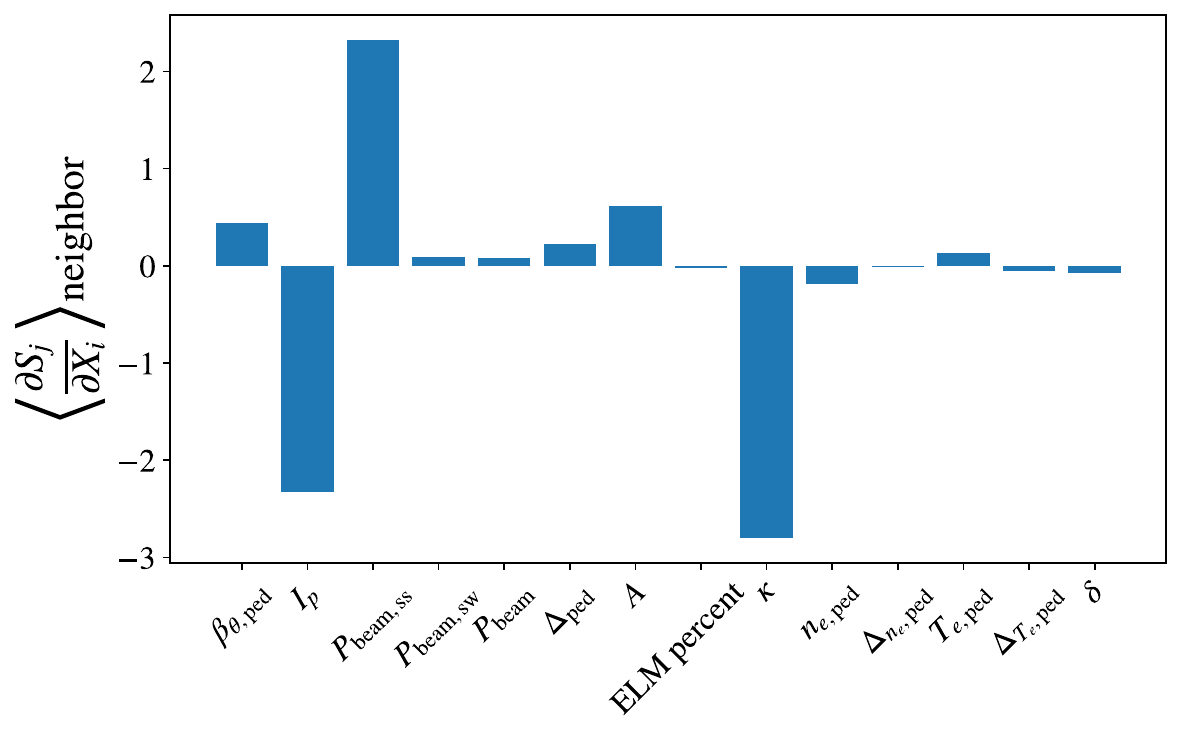}
    \caption{Case C}
    \end{subfigure}
    \caption{Sensitivity of MCDM score to input parameters in the vicinity of the Pareto maximum for Cases A and C. See \Cref{eq:mcdm_derivative} and surrounding discussion for more details.}
    \label{fig:nearest_neighbor_average}
\end{figure*}

As a short demonstration of how HIPED could be used to enhance experiments beyond their current performance, we show the derivative of the MCDM score $S_j$ with respect to input parameters. This derivative is evaluated in the vicinity of the MCDM optimal point -- this is obtained by averaging over the 50 points with scores closest to the MCDM optimal point,
\begin{equation}
\left\langle \frac{\partial S_j}{\partial X_i} \right\rangle_\mathrm{neighbor},
\label{eq:mcdm_derivative}
\end{equation}
where $X_i$ is an input parameter normalized between 0 and 1, and $\langle \rangle_\mathrm{neighbor}$ indicates the average over the 50 nearest neighbours. The results are shown in \Cref{fig:nearest_neighbor_average} -- for Case A [\Cref{fig:nearest_neighbor_average}(a)], the two most important parameters for improving performance are $I_p$ and $\kappa$: increasing both of them in the vicinity of the MCDM optimal point increases lower ELM frequency performance. In contrast, for Case C [\Cref{fig:nearest_neighbor_average}(b)], both $I_p$ and $\kappa$ must be decreased to increase high $\beta_N$ performance -- for Case C, increasing the south beam power $P_\mathrm{beam,ss}$ is also important to increase performance. This presents an approach to further improving the best discharges on an experiment.

There are many directions for future application. For example, while one of our Pareto optimization parameters is $T_\mathrm{ELM}$, this does not include information about the energy and particles expelled by the ELM. Other crucial aspects of ELM-free operation such as impurity transport \cite{Putterich2011} could also be included.

\section{Discussion}
\label{sec:discussion}

We have introduced HIPED, a new framework for predicting and optimizing core and pedestal performance in spherical tokamaks using a large MAST-U database. Our main findings include:  
\begin{enumerate}
\item \emph{Core-edge coupling.} Simple power-law relationships, such as $\Delta_{\mathrm{ped}}\sim\sqrt{\beta_{\theta,\mathrm{ped}}}$, are not suitable in our dataset. Additional parameters like $\beta_N$, $f_G$, $\kappa$, and $P_{\mathrm{beam}}$ significantly improve accuracy.
\item \emph{Random Forest performance.} The RF models achieve robust accuracy ($R^2 \approx 0.7$--0.8) and outperform simpler regressions. Feature-importance [e.g. \Cref{fig:betapedtheta_RF}] and SHAP analyses [e.g. \Cref{fig:shap_RF_Model_betaped}] clarify which parameters dominate.
\item \emph{Multi-objective optimization.} We use Pareto methods to identify high-value discharges and detail the control room parameters required to access them. This multi-objective approach helps design experiments with complex performance targets. As demonstrated in this paper, this functionality can be extended beyond just pedestal parameters, and can be used for optimization on other tokamak machines. We presented an example of how this framework can show how to change the most important parameters in the vicinity of high performance discharges [\Cref{fig:nearest_neighbor_average}].
\end{enumerate}

Future extensions include deeper statistical analysis of the multi-parameter optimization, extending HIPED analysis to more MAST-U campaigns and other machines, real-time implementations for experimental feedback, and broader integration with first-principles codes. Additionally, given the range of transport candidates in the pedestal, the relatively high time resolution of the fitted pedestal profiles could constrain the particle, momentum, and heat transport mechanisms in the pedestal \cite{Hubbard2000,Chang2004,Angioni_2009,Callen2010,Hatch2017,Kotschenreuther2019,Guttenfelder2021,Chapman2022,Belli_2023,Hatch2024}. Since the pedestal couples the core and scrape-off layer (SOL), including SOL and divertor quantities would allow for further optimization of core-pedestal-SOL coupling \cite{Neuhauser2002,Lipschultz2007,Luda2020,Wilks2021,Stagni2022,Snyder2024,Zhang2024,RodriguezFernandez_2024,Eich2025,Welsh2025}. In addition, further expansions to the feature set—e.g., including fast-ion physics or edge diagnostic inputs—may improve predictions of pedestal parameters.

\section*{Acknowledgments}

We gratefully acknowledge the contributions of the MAST-U team in producing the dataset. This work was supported by the U.S. Department of Energy under Contracts DE-AC02-09CH11466, DE-SC0022270, DE-SC0022272. This work was also made possible by funding from the U.S. Department of Energy for the Summer Undergraduate Laboratory Internship (SULI) program. The United States Government retains a non-exclusive, paid-up, irrevocable, world-wide license to publish or reproduce the published form of this manuscript, or allow others to do so, for United States Government purposes.

This work has received support from EPSRC Grant EP/T012250/1 and by the EPSRC Energy Programme Grant EP/W006839/1. This work has been carried out within the framework of the EUROfusion Consortium, partially funded by the European Union via the Euratom Research and Training Programme (Grant Agreement No. 101052200 – EUROfusion). Views and opinions expressed are however those of the author(s) only and do not necessarily reflect those of the European Union or the European Commission. Neither the European Union nor the European Commission can be held responsible for them.

\section{\label{s:Data Availability}Data Availability}
The data supporting this study are available from the corresponding author upon reasonable request. Please contact \texttt{PublicationsManager@ukaea.uk} for further information on MAST-U data. An active GitLab repository can be accessed at https://git.pppl.gov/jclark/pypedestaldatabase which contains the pedestal database filtering function and tools. HIPED will be made available in an open-source repository following publication of this work.

\begin{table*}
\caption{Key quantities used in this work. $\langle \rangle$ indicates a plasma volume average.}
\begin{ruledtabular}
\centering
\begin{tabular}{ ccccc  }
Name & Quantity & Units & Definition & Physics/Control Room Param.  \\
\hline
Normalized pedestal height & $\beta_{\theta,\mathrm{ped}}$ & - &  
$2\,p_{e,\mathrm{ped}} / \bigl(\mu_0\,I_p / l \bigr)^2$ & Physics \\
Electron pressure pedestal height & $p_{e,\mathrm{ped}}$ & kPa &  
$k_B\,n_{e,\mathrm{ped}}\,T_{e,\mathrm{ped}}$ & Physics \\
Electron density pedestal height & $n_{e,\mathrm{ped}}$ & $10^{20}$\,m$^{-3}$ & & Physics \\
Electron temperature pedestal height & $T_{e,\mathrm{ped}}$ & keV & & Physics \\
Total pedestal width & $\Delta_{\mathrm{ped}}$ & (norm.\ flux) &  
$( \Delta_{n_e,\mathrm{ped}} + \Delta_{T_e,\mathrm{ped}} )/2$ & Physics \\
Electron density pedestal width & $\Delta_{n_e,\mathrm{ped}}$ & (norm.\ flux) & & Physics \\
Electron temperature pedestal width & $\Delta_{T_e,\mathrm{ped}}$ & (norm.\ flux) & & Physics \\
Normalized plasma pressure & $\beta_N$ & - & $( 2\,\mu_0\,\langle p\rangle / B_{T,0}^2 )\,\tfrac{a B_{T,0}}{I_p}$ & Physics \\
Total plasma current & $I_p$ & MA & & Control Room \\
Greenwald fraction & $f_G$ & - & $\langle n_e\rangle / (I_p/(\pi a^2))$ & Control Room \\
Plasma elongation & $\kappa$ & - & & Control Room \\
Plasma triangularity & $\delta$ & - & & Control Room \\
Aspect ratio & $A$ & - & & Control Room \\
Total NBI power & $P_{\mathrm{beam}}$ & MW & & Control Room \\
South NBI power & $P_{\mathrm{beam,ss}}$ & MW & & Control Room \\
Southwest NBI power & $P_{\mathrm{beam,sw}}$ & MW & & Control Room \\
\end{tabular}
\end{ruledtabular}
\label{tab:tab0}
\end{table*}

\appendix

\section{Random Forest Model Details}
\label{app:rfdetails}

Here we summarize training details for the Random Forest (RF) models. Each RF uses 100--300 trees with a maximum depth of 30 to mitigate overfitting. Bootstrap aggregation (bagging) trains each tree on a random subset. We compute feature importance by Gini impurity or permutation-based metrics, yielding consistent results. Sensitivity checks of various model fitting parameters suggest these defaults suffice to achieve stable performance on our MAST-U dataset.

\section{Pareto-Optimal Trajectories}
\label{app:pareto_optimal}

\begin{figure*}[bt!]
    \begin{subfigure}[t]{0.24\textwidth}
    \includegraphics[width=\textwidth]{5NN_unique_trajectory_in_fG_betaN_a_opt4.pdf}
    \caption{Case A}
    \end{subfigure}
    \begin{subfigure}[t]{0.24\textwidth}
    \includegraphics[width=\textwidth]{5NN_unique_trajectory_in_ne_TELM_a_opt4.pdf}
    \caption{Case A}
    \end{subfigure}
    \begin{subfigure}[t]{0.24\textwidth}
    \includegraphics[width=\textwidth]{5NN_unique_trajectory_in_Deltaped_betaped_a_opt4.pdf}
    \caption{Case A}
    \end{subfigure}
    \begin{subfigure}[t]{0.24\textwidth}
    \includegraphics[width=\textwidth]{5NN_unique_trajectory_in_dens_temp_heights_a_opt4.pdf}
    \caption{Case A}
    \end{subfigure}
    \begin{subfigure}[t]{0.24\textwidth}
    \includegraphics[width=\textwidth]{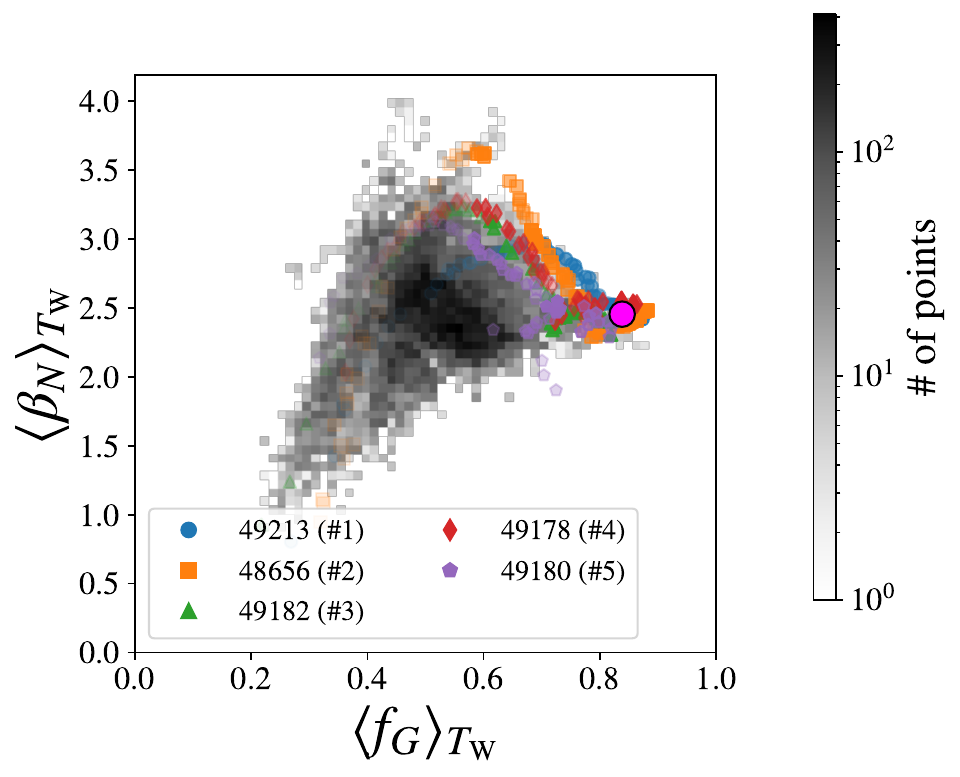}
    \caption{Case B}
    \end{subfigure}
    \begin{subfigure}[t]{0.24\textwidth}
    \includegraphics[width=\textwidth]{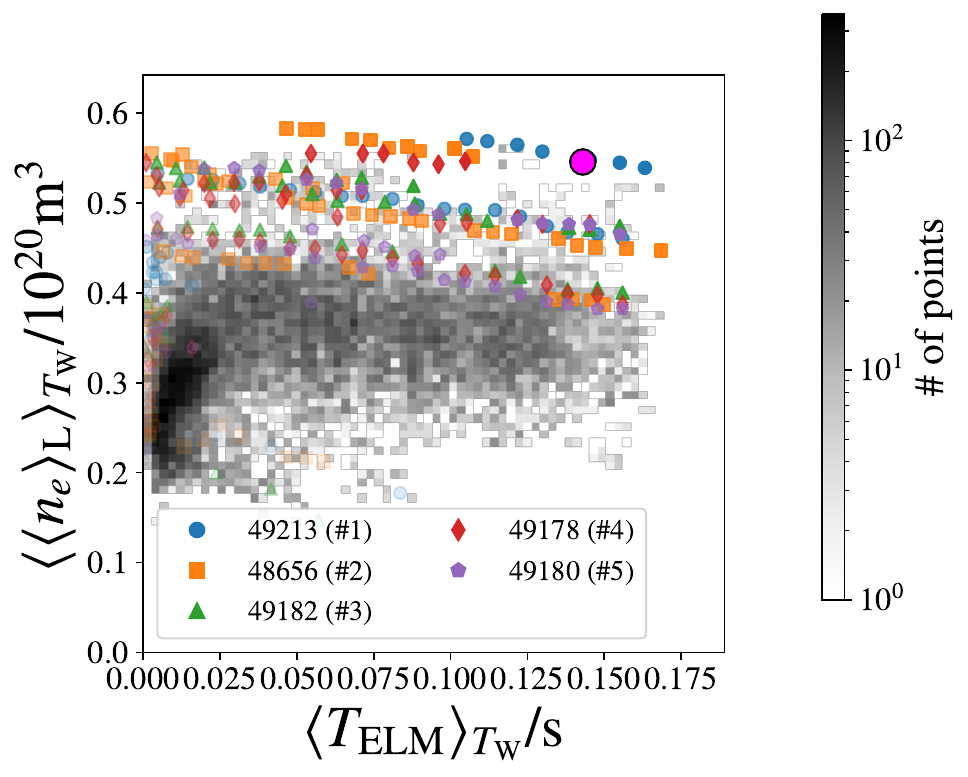}
    \caption{Case B}
    \end{subfigure}
    \begin{subfigure}[t]{0.24\textwidth}
    \includegraphics[width=\textwidth]{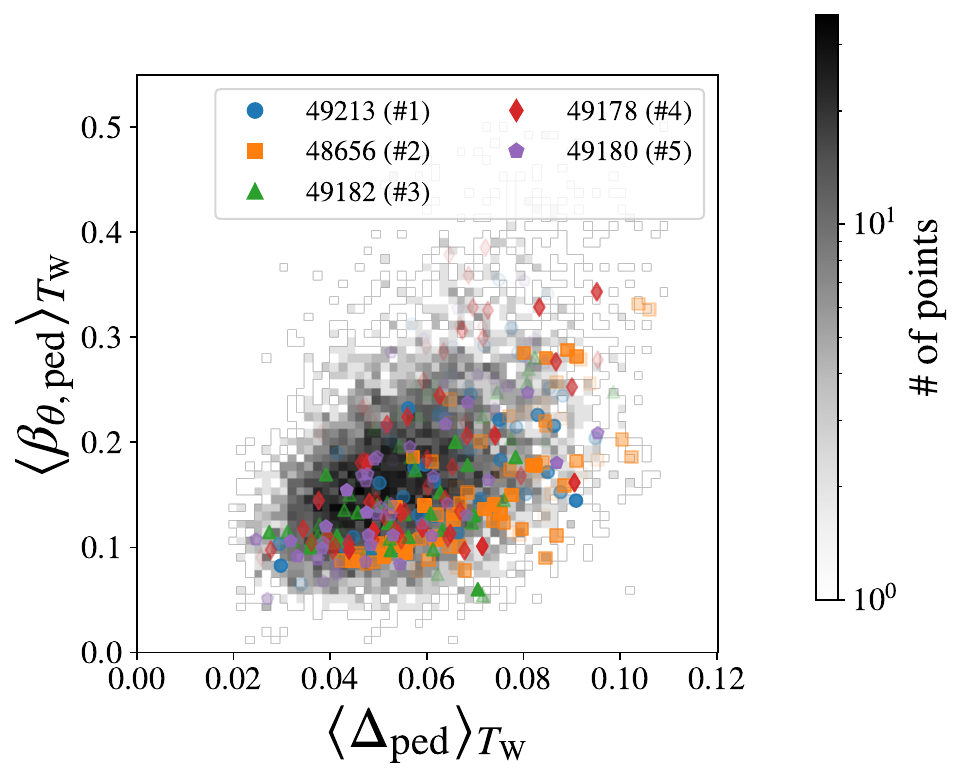}
    \caption{Case B}
    \end{subfigure}
    \begin{subfigure}[t]{0.24\textwidth}
    \includegraphics[width=\textwidth]{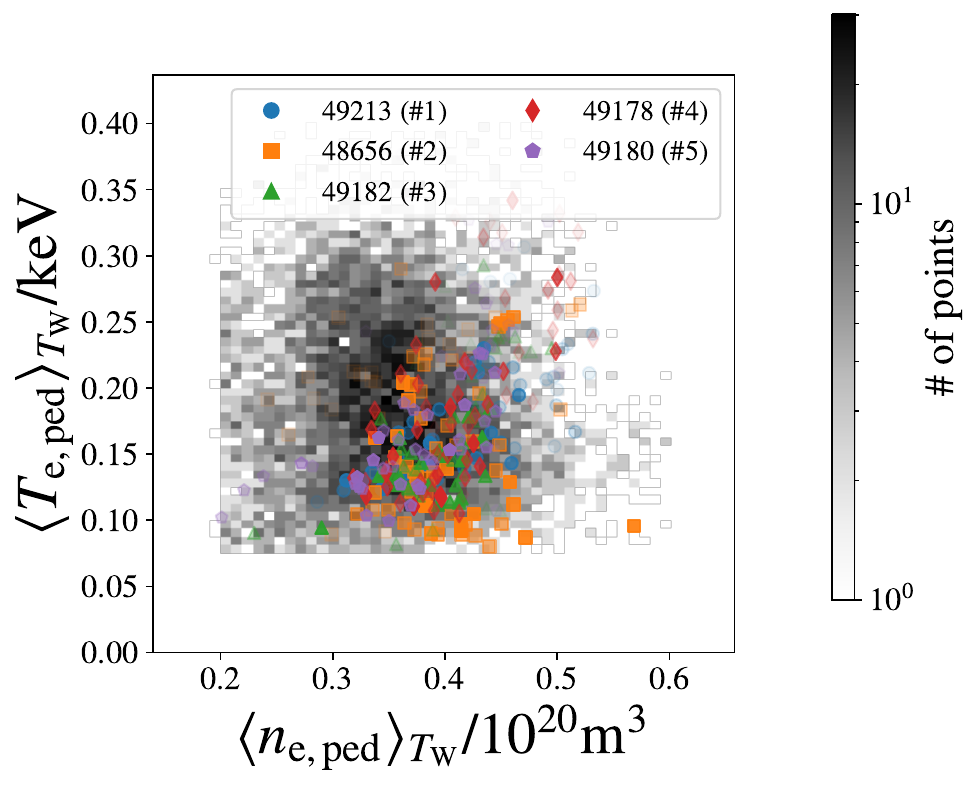}
    \caption{Case B}
    \end{subfigure}
    \begin{subfigure}[t]{0.24\textwidth}
    \includegraphics[width=\textwidth]{5NN_unique_trajectory_in_fG_betaN_c_opt4.pdf}
    \caption{Case C}
    \end{subfigure}
    \begin{subfigure}[t]{0.24\textwidth}
    \includegraphics[width=\textwidth]{5NN_unique_trajectory_in_ne_TELM_c_opt4.pdf}
    \caption{Case C}
    \end{subfigure}
    \begin{subfigure}[t]{0.24\textwidth}
    \includegraphics[width=\textwidth]{5NN_unique_trajectory_in_Deltaped_betaped_c_opt4.pdf}
    \caption{Case C}
    \end{subfigure}
    \begin{subfigure}[t]{0.24\textwidth}
    \includegraphics[width=\textwidth]{5NN_unique_trajectory_in_dens_temp_heights_c_opt4.pdf}
    \caption{Case C}
    \end{subfigure}
    \begin{subfigure}[t]{0.24\textwidth}
    \includegraphics[width=\textwidth]{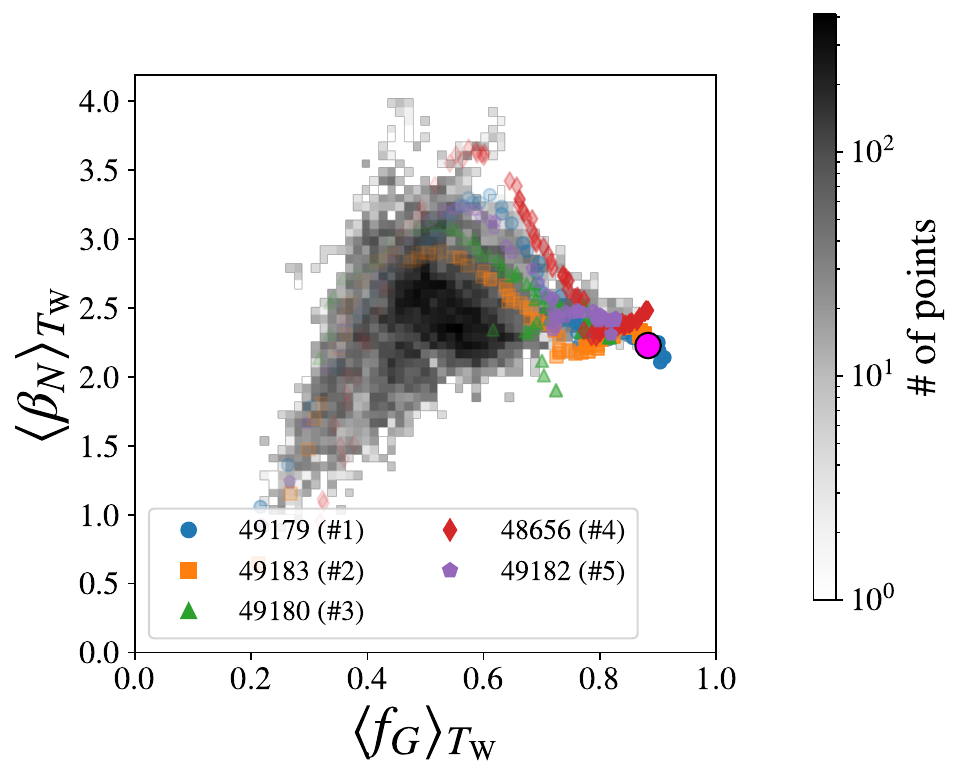}
    \caption{Case D}
    \end{subfigure}
    \begin{subfigure}[t]{0.24\textwidth}
    \includegraphics[width=\textwidth]{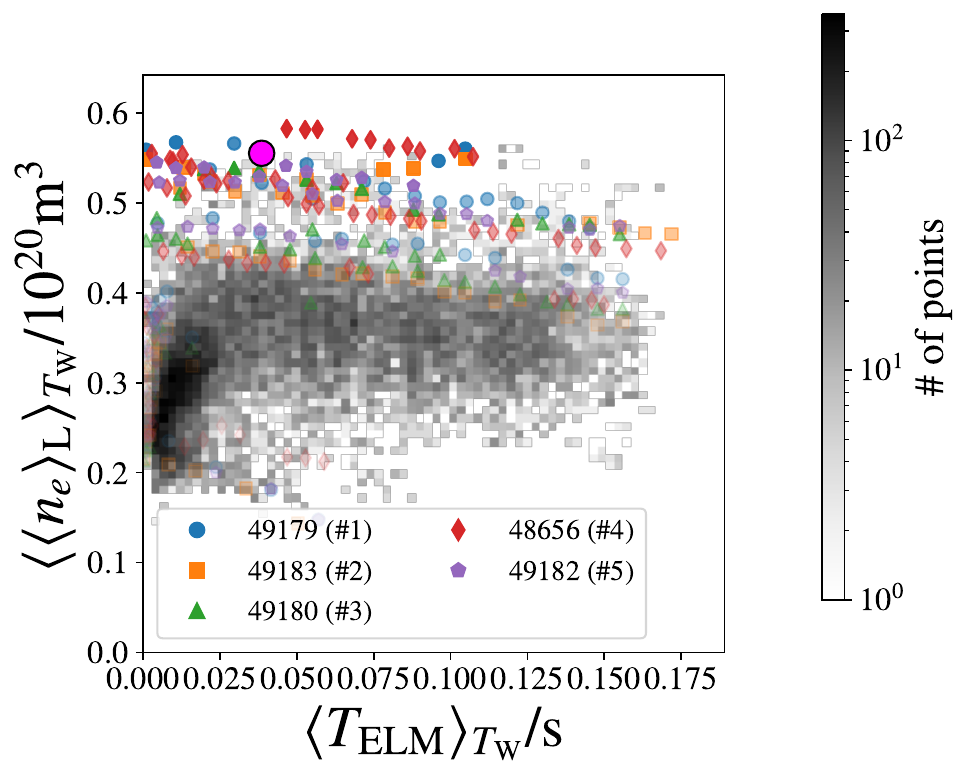}
    \caption{Case D}
    \end{subfigure}
    \begin{subfigure}[t]{0.24\textwidth}
    \includegraphics[width=\textwidth]{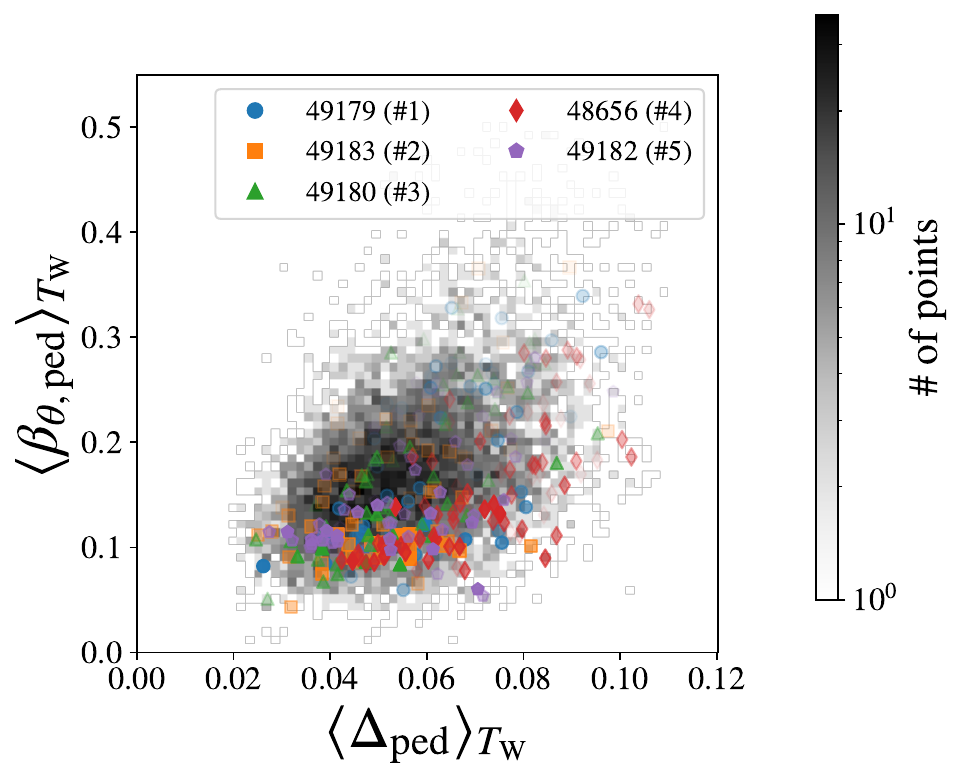}
    \caption{Case D}
    \end{subfigure}
    \begin{subfigure}[t]{0.24\textwidth}
    \includegraphics[width=\textwidth]{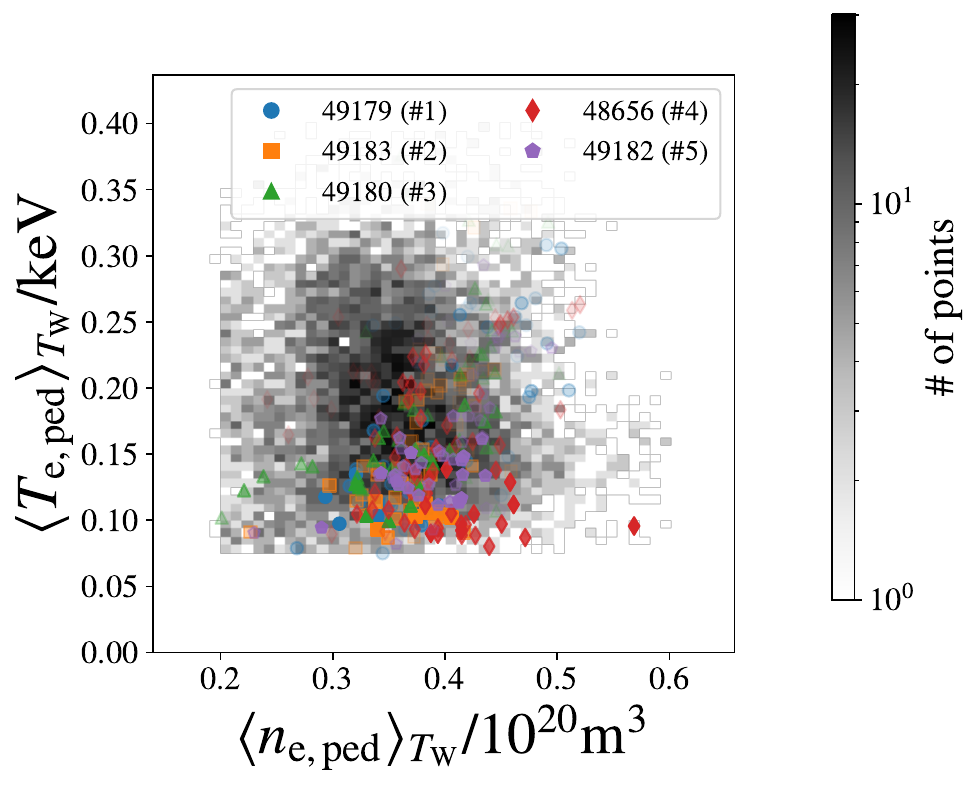}
    \caption{Case D}
    \end{subfigure}
    \caption{Physics parameters for Pareto-optimal time points in MAST-U, Cases A-D (rows 1 -4) and quantities $(\beta_N,f_G)$, $(\langle n_e \rangle_L ,  T_\mathrm{ELM}   )$, $(\beta_{\theta, \mathrm{ped} },\Delta_\mathrm{ped})$, $(T_{\mathrm{e,ped}},n_{\mathrm{e,ped}})$ (columns 1 - 4).}
    \label{fig:pareto_further_performance}
\end{figure*}

\begin{figure*}[bt!]
    \begin{subfigure}[t]{0.24\textwidth}
    \includegraphics[width=\textwidth]{5NN_unique_trajectory_in_fG_Ip_a_opt4.pdf}
    \caption{Case A}
    \end{subfigure}
    \begin{subfigure}[t]{0.24\textwidth}
    \includegraphics[width=\textwidth]{5NN_unique_trajectory_in_triangularity_elongation_a_opt4.pdf}
    \caption{Case A}
    \end{subfigure}
    \begin{subfigure}[t]{0.24\textwidth}
    \includegraphics[width=\textwidth]{5NN_unique_trajectory_in_NBIss_NBIsw_a_opt4.pdf}
    \caption{Case A}
    \end{subfigure}
    \begin{subfigure}[t]{0.24\textwidth}
    \includegraphics[width=\textwidth]{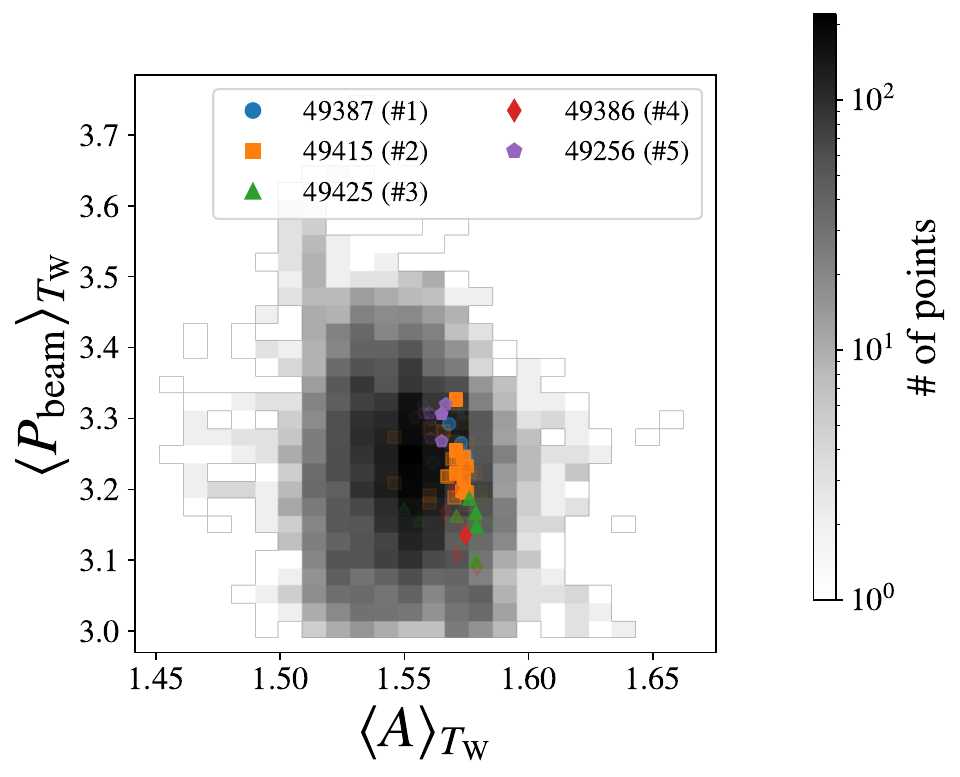}
    \caption{Case A}
    \end{subfigure}
    \begin{subfigure}[t]{0.24\textwidth}
    \includegraphics[width=\textwidth]{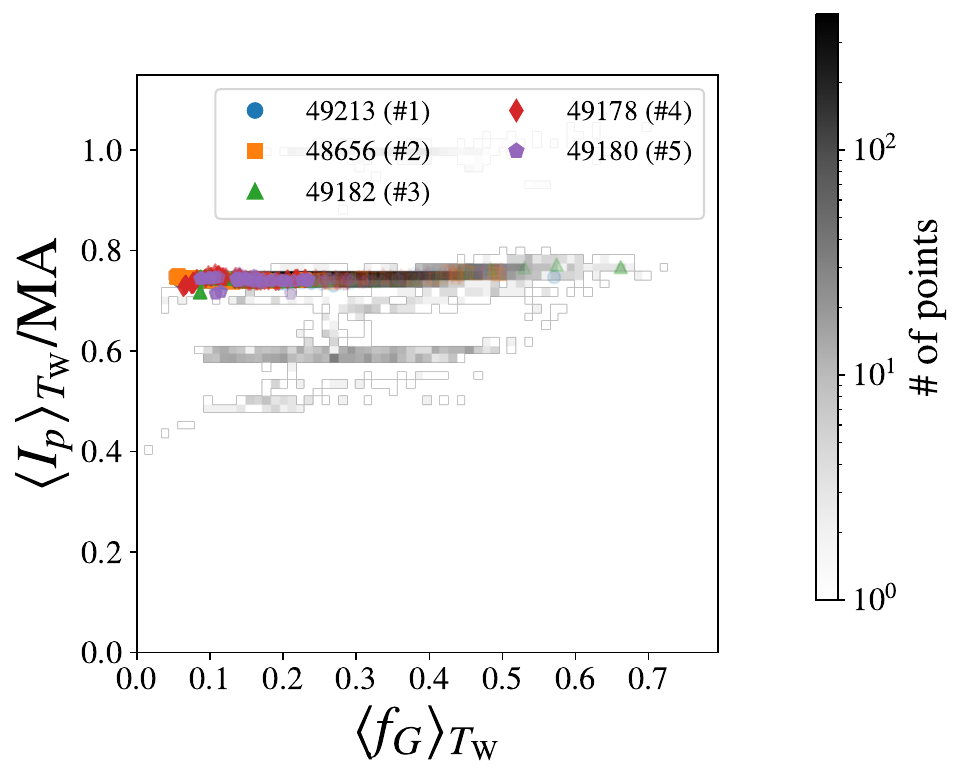}
    \caption{Case B}
    \end{subfigure}
    \begin{subfigure}[t]{0.24\textwidth}
    \includegraphics[width=\textwidth]{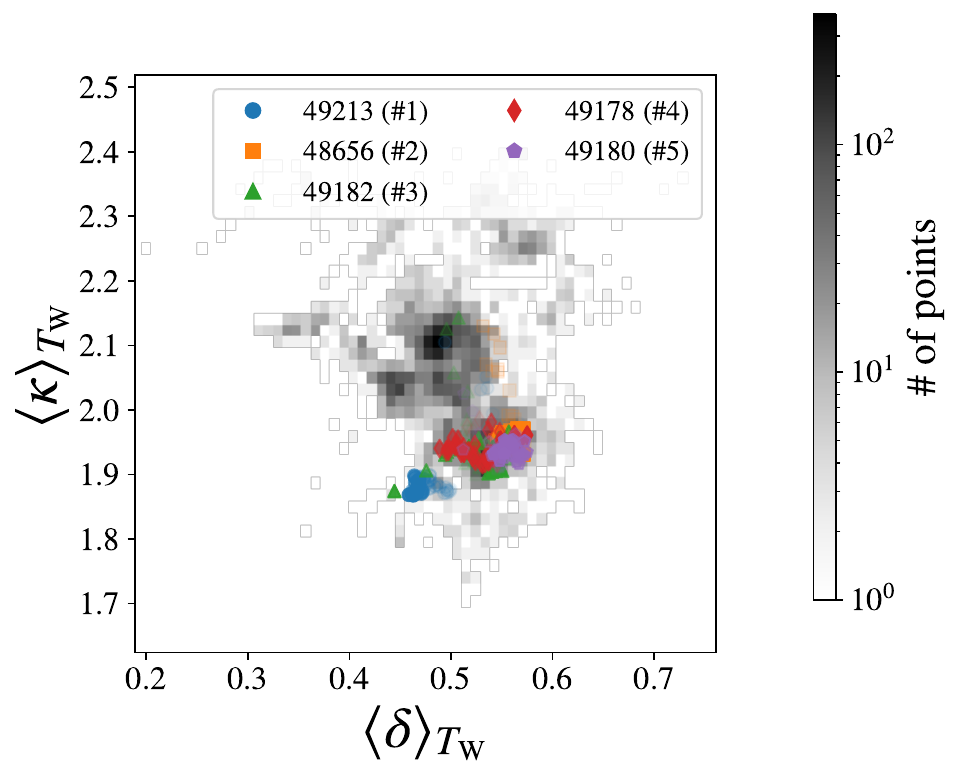}
    \caption{Case B}
    \end{subfigure}
    \begin{subfigure}[t]{0.24\textwidth}
    \includegraphics[width=\textwidth]{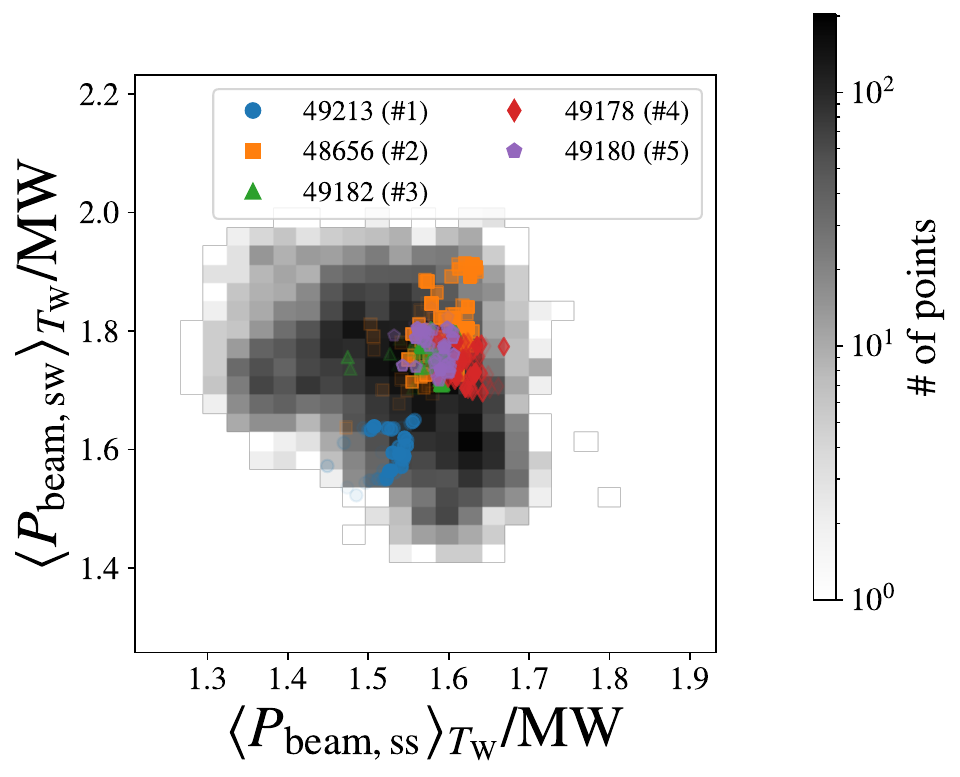}
    \caption{Case B}
    \end{subfigure}
    \begin{subfigure}[t]{0.24\textwidth}
    \includegraphics[width=\textwidth]{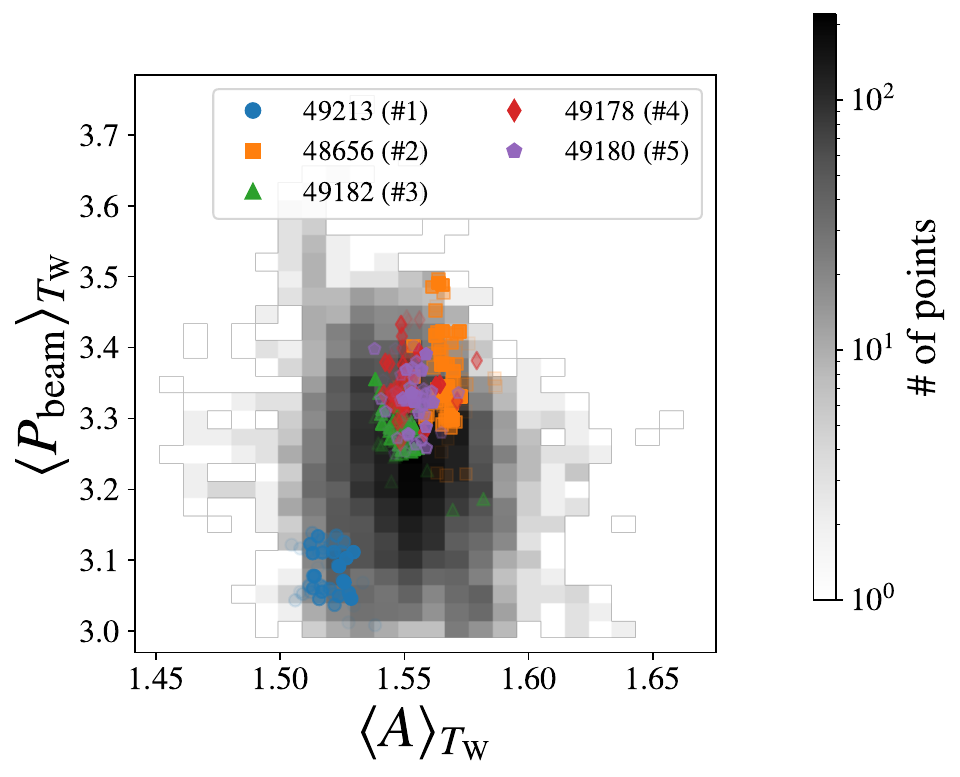}
    \caption{Case B}
    \end{subfigure}
    \begin{subfigure}[t]{0.24\textwidth}
    \includegraphics[width=\textwidth]{5NN_unique_trajectory_in_fG_Ip_c_opt4.pdf}
    \caption{Case C}
    \end{subfigure}
    \begin{subfigure}[t]{0.24\textwidth}
    \includegraphics[width=\textwidth]{5NN_unique_trajectory_in_triangularity_elongation_c_opt4.pdf}
    \caption{Case C}
    \end{subfigure}
    \begin{subfigure}[t]{0.24\textwidth}
    \includegraphics[width=\textwidth]{5NN_unique_trajectory_in_NBIss_NBIsw_c_opt4.pdf}
    \caption{Case C}
    \end{subfigure}
    \begin{subfigure}[t]{0.24\textwidth}
    \includegraphics[width=\textwidth]{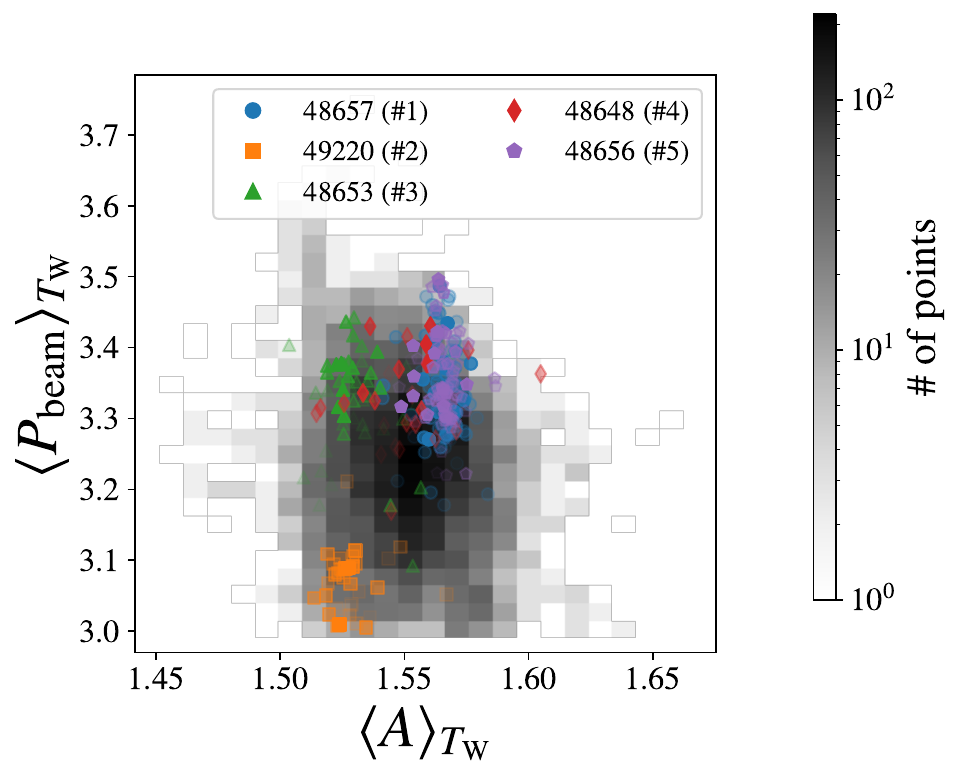}
    \caption{Case C}
    \end{subfigure}
    \begin{subfigure}[t]{0.24\textwidth}
    \includegraphics[width=\textwidth]{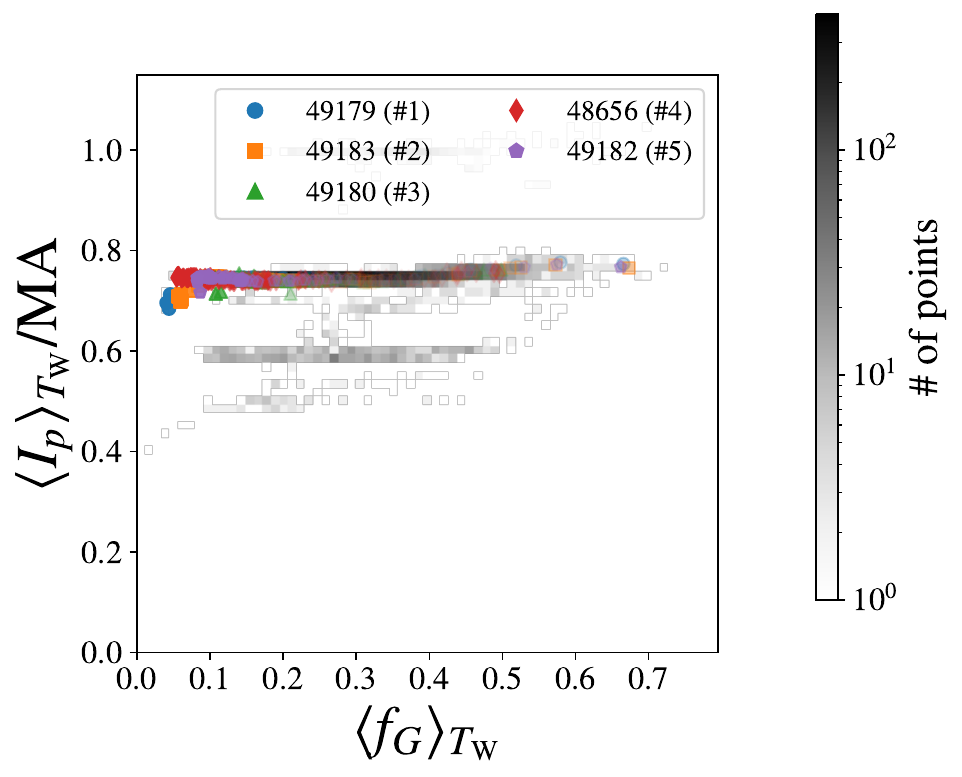}
    \caption{Case D}
    \end{subfigure}
    \begin{subfigure}[t]{0.24\textwidth}
    \includegraphics[width=\textwidth]{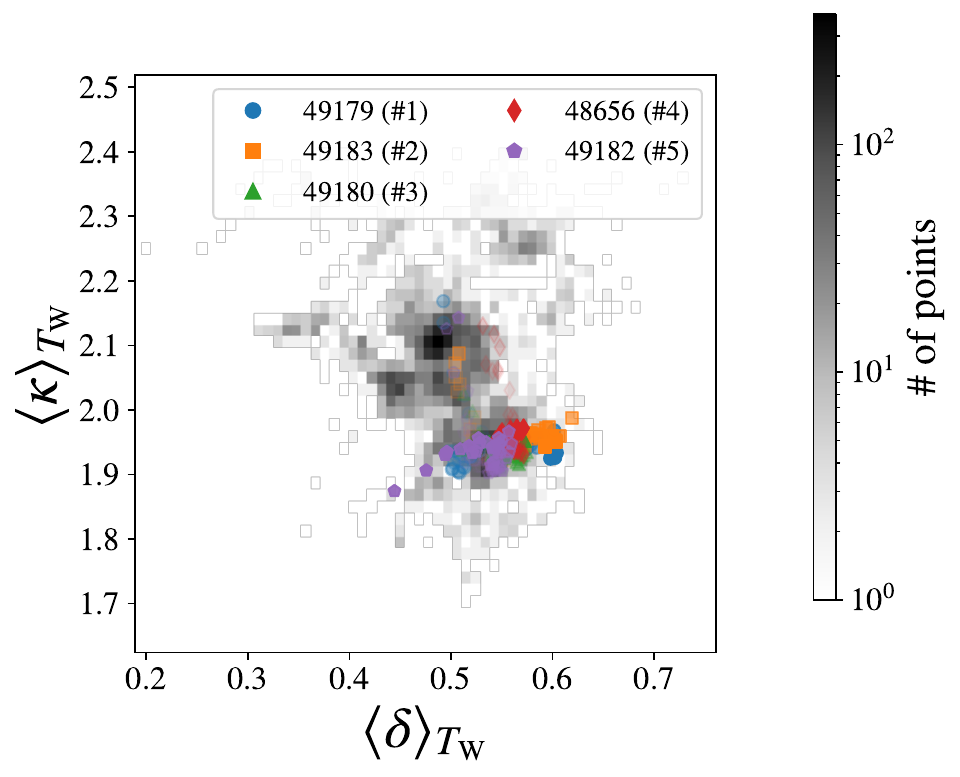}
    \caption{Case D}
    \end{subfigure}
    \begin{subfigure}[t]{0.24\textwidth}
    \includegraphics[width=\textwidth]{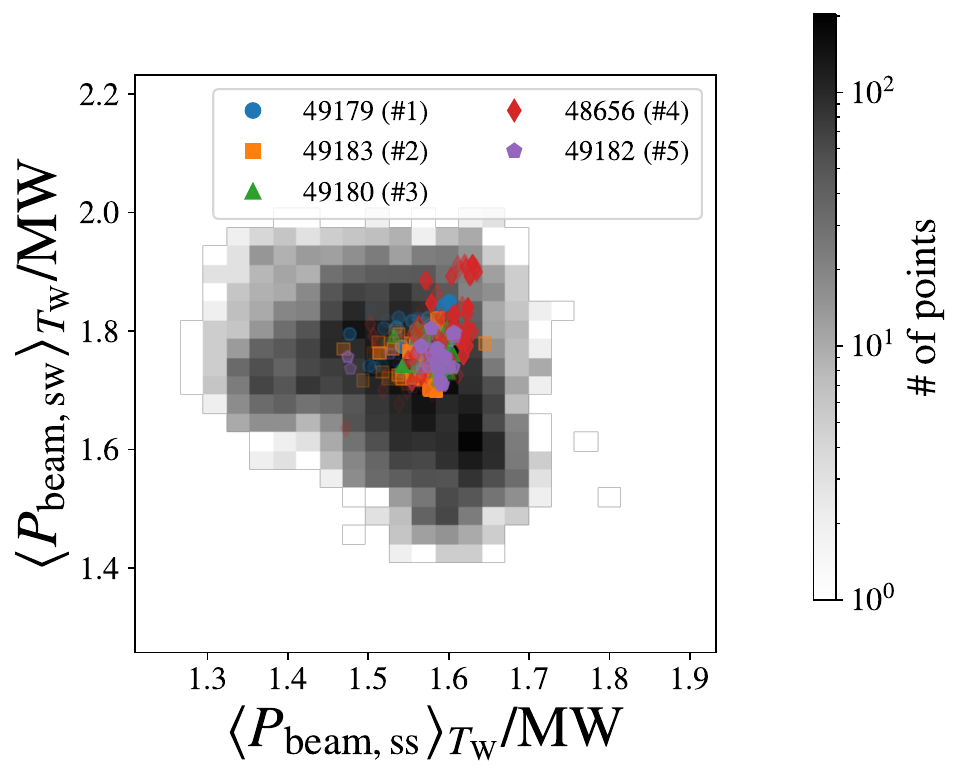}
    \caption{Case D}
    \end{subfigure}
    \begin{subfigure}[t]{0.24\textwidth}
    \includegraphics[width=\textwidth]{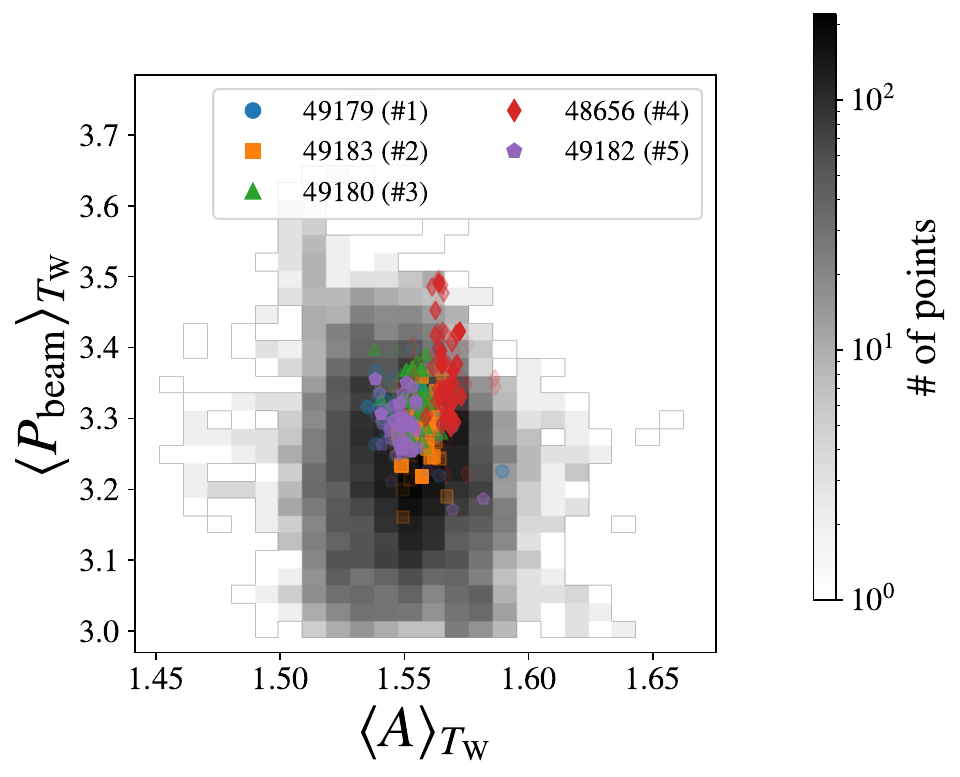}
    \caption{Case D}
    \end{subfigure}
    \caption{Control-room parameters for Pareto-optimal time points in MAST-U, Cases A- D (rows 1 - 4) and quantities $(I_p,f_G)$, $(\delta,\kappa)$, $(A,P_\mathrm{beam})$, $(P_{\mathrm{beam,ss}},P_{\mathrm{beam,sw}})$ (columns 1 - 4).}
    \label{fig:pareto_further_control_room}
\end{figure*}

We provide additional detail on Pareto-optimal MAST-U trajectories under the weighting Cases A-D (see \Cref{table:MASTU_MCDMweightings} in \Cref{sec:Pareto_optimization}). \Cref{fig:pareto_further_performance} shows core and pedestal performance parameters ($n_{e,\mathrm{ped}}$, $T_{e,\mathrm{ped}}$, $\Delta_{n_{e,\mathrm{ped}}}$, $\Delta_{T_{e,\mathrm{ped}}}$) among shots closest to the Pareto fronts.\Cref{fig:pareto_further_control_room} shows the control room parameters for Cases A - D.  As the $T_\mathrm{ELM}$ emphasis decreases, $T_{e,\mathrm{ped}}$ tends to increase but $n_{e,\mathrm{ped}}$ decreases.

\section{MCDM Details}
\label{app:MCDM}

\begin{figure}[bt!]
    \centering
    \begin{subfigure}[t]{0.48\textwidth}
    \centering
    \includegraphics[width=1.0\textwidth]{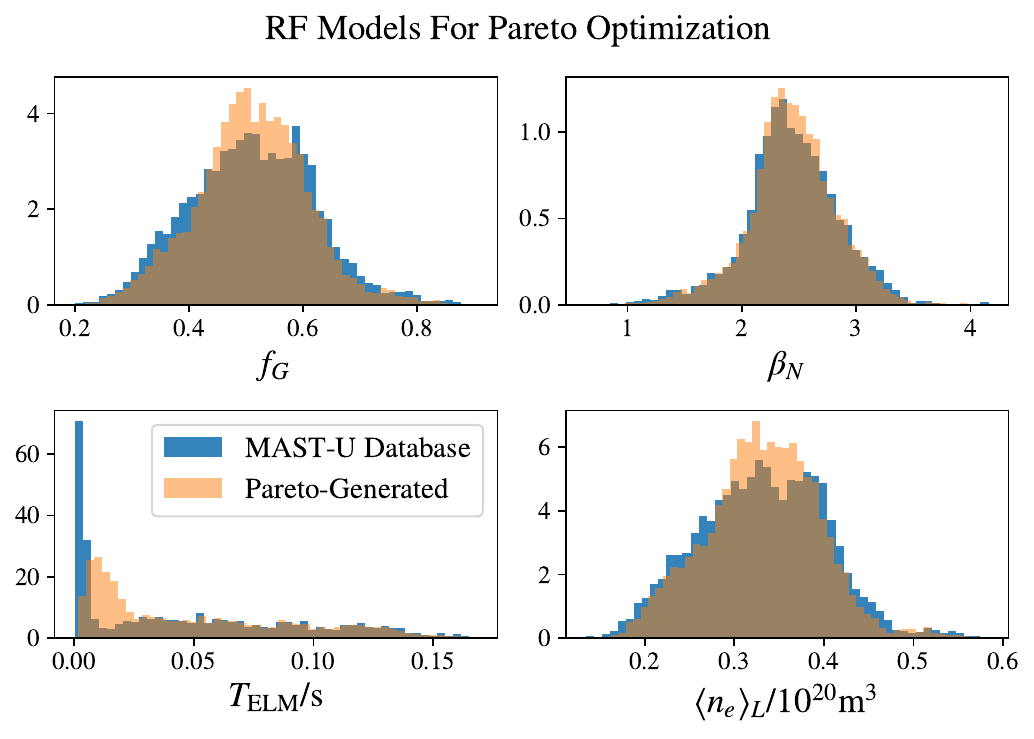}
    \end{subfigure}
    \caption{Random Forest model performance on $f_G$, $\beta_N$, $T_\mathrm{ELM}$, and $\langle n_e \rangle_L$. Y-axis measures the count number.}
    \label{fig:pareto_ML_models}
\end{figure}

We now show how well the model captures the underlying data for $f_G$, $\beta_N$, $T_\mathrm{ELM}$, and $\langle n_e \rangle_L$. Capturing the behavior of $T_\mathrm{ELM}$ with an RF model is challenging because of the distribution being highly skewed towards very small values of $T_\mathrm{ELM}$. To that end, before training an RF model for $T_\mathrm{ELM}$ we oversample very small and large values in the RF training. The RF training for $f_G$, $\beta_N$, and $\langle n_e \rangle_L$ is straightforward, not requiring any oversampling or hypertuning. In \Cref{fig:pareto_ML_models}, we compare the RF models for $f_G$, $\beta_N$, $T_\mathrm{ELM}$, and $\langle n_e \rangle_L$ in the MAST-U database with their predictions. Notably, the model struggles at very low values of $T_\mathrm{ELM}$. However, we justify this approximation because for Pareto optimization, we are mainly interested in discharges with larger $T_\mathrm{ELM}$ values -- but with the caveat that the Pareto fronts may have some errors with very small $T_\mathrm{ELM}$ values.

\section{RF Model Without $\beta_N$} \label{app:noBetaN}

\begin{figure}[bt!]
    \centering
    \begin{subfigure}[t]{0.48\textwidth}
    \centering
    \includegraphics[width=1.0\textwidth]{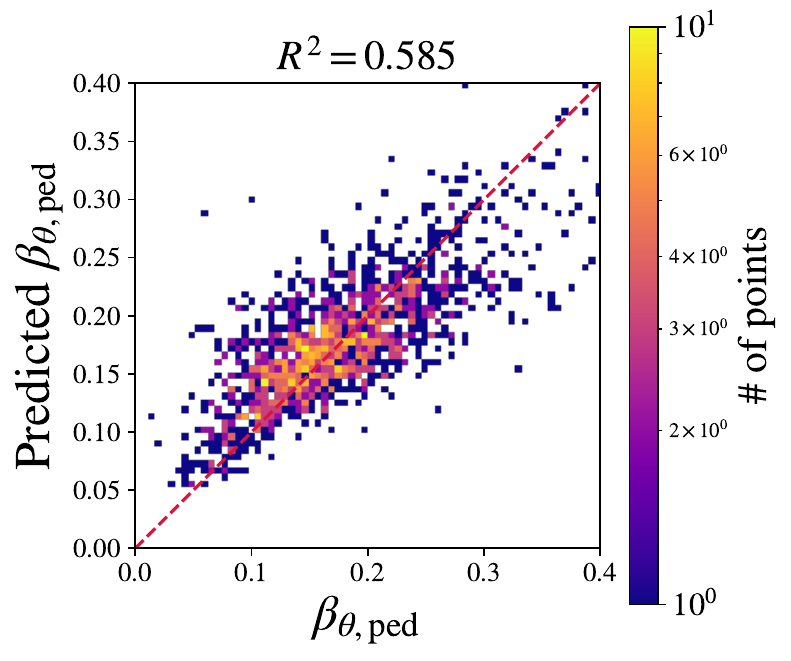}
    \end{subfigure}
    \centering
    \begin{subfigure}[t]{0.48\textwidth}
    \centering
    \includegraphics[width=1.0\textwidth]{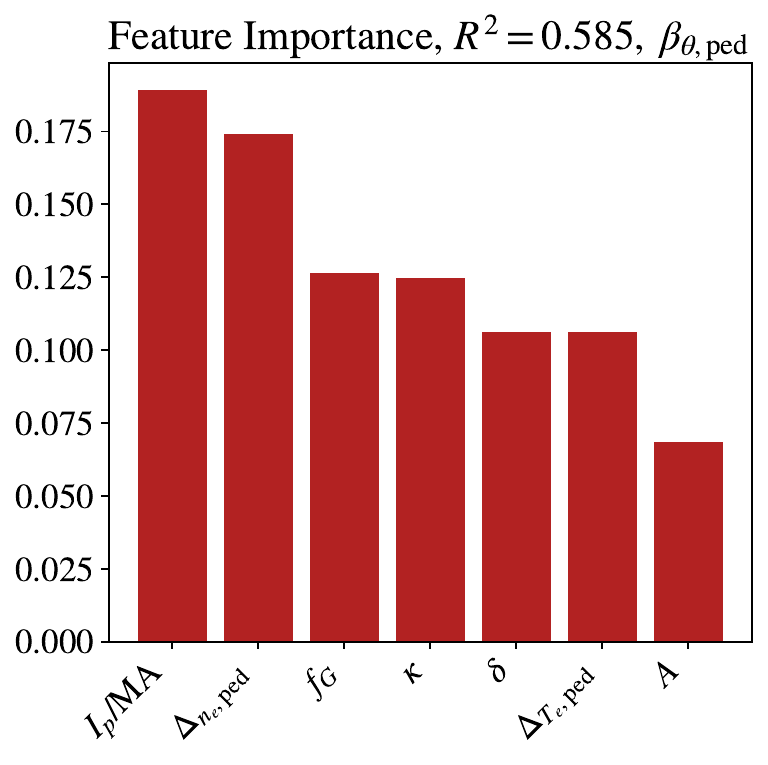}
    \end{subfigure}
    \caption{Random Forest model for $\beta_{\theta,\mathrm{ped} }$ excluding $\beta_N$ as an input parameter. Compare with \Cref{fig:betapedtheta_RF}(a) and (d) where $\beta_N$ is included as an input parameter.}
    \label{fig:no_betaN}
\end{figure}

For completeness, in this section we show the effect of excluding $\beta_N$ on the accuracy of an RF model trained for predicting $\beta_{\theta, \mathrm{ped}}$. Excluding $\beta_N$ in the fitting parameters is motivated by the fact that the positive relation between $\beta_{\theta, \mathrm{ped} }$ and $\beta_N$ might be obvious. The RF model fit and feature importances are shown in \Cref{fig:no_betaN}(a) and (b). \Cref{fig:no_betaN}(a) shows that $R^2 = 0.59$, which is a significant decrease from the RF model in \Cref{fig:betapedtheta_RF}(a) that achieved $R^2 = 0.76$. The feature importance rankings in \Cref{fig:no_betaN}(b) are very different to the RF model that included $\beta_{\theta,\mathrm{ped} }$ [\Cref{fig:betapedtheta_RF}(d)]. When $\beta_N$ is excluded, the most important input parameter is $I_p$, whereas this $I_p$ was not particularly important in the RF model including $\beta_N$. This difference in the importance of $I_p$ is not too surprising given that $I_p$ enters the definition of $\beta_N$ -- a model including $\beta_N$ captures some of the behavior of $I_p$. Furthermore, the definition of $\beta_{\theta, \mathrm{ped} }$ with the scalings $B_\mathrm{pol} \sim I_p^2$ and $p_{e,\mathrm{ped} } \sim I_p$ \cite{Cordey1991} suggests the scaling $\beta_{\theta, \mathrm{ped} } \sim 1 / I_p$ at fixed flux-surface circumference. Therefore, the strong dependence of $\beta_{\theta, \mathrm{ped} }$ on $I_p$ in the RF models is not surprising.

\section{SHAP for $n_{e,\mathrm{ped} }$ and $T_{e,\mathrm{ped} }$} \label{app:nepedTepedSHAP}

\begin{figure*}[!tb]
    \centering
    \begin{subfigure}[t]{0.73\textwidth}
    \includegraphics[width=\textwidth]{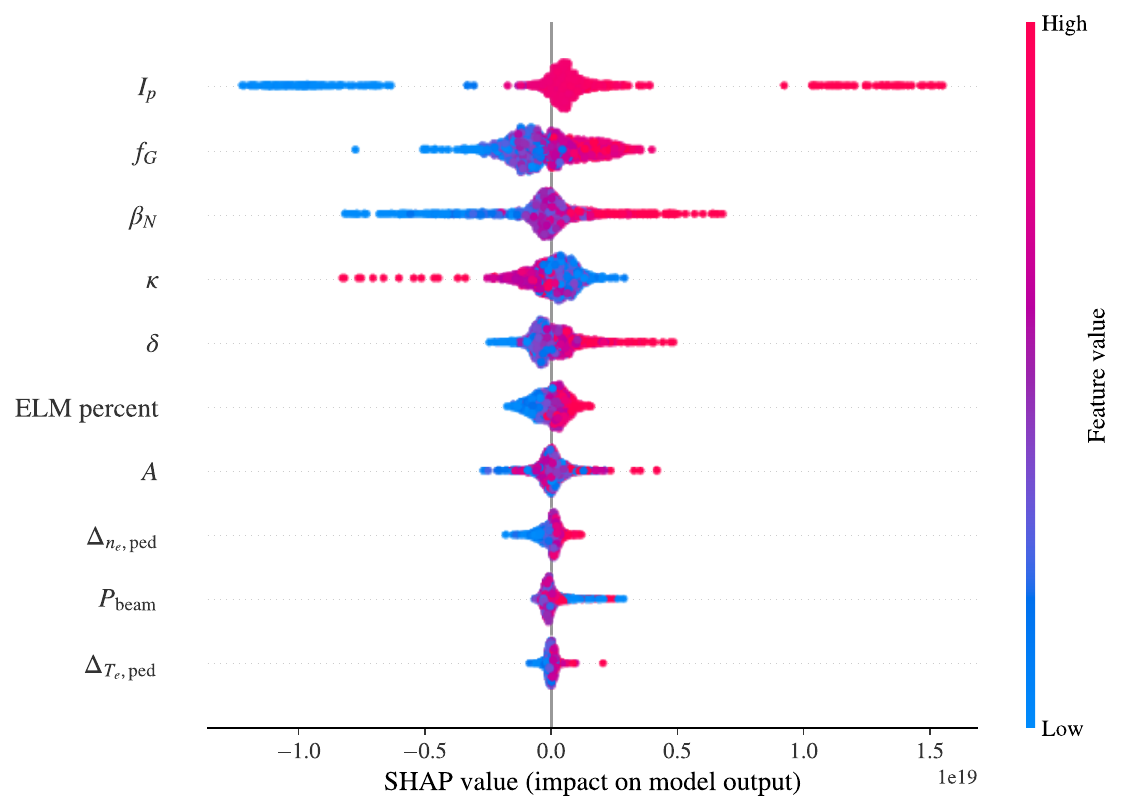}
    \caption{$n_{e,\mathrm{ped}}$}
    \end{subfigure}
    \begin{subfigure}[t]{0.73\textwidth}
    \includegraphics[width=\textwidth]{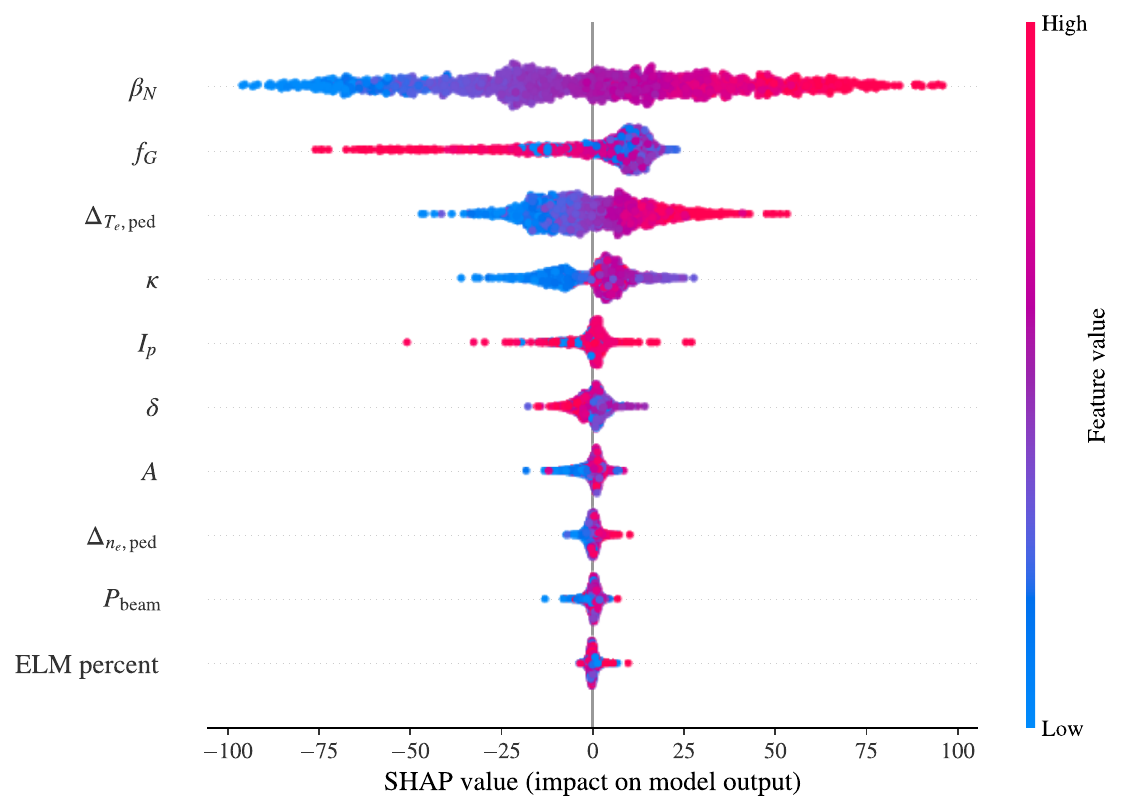}
    \caption{$T_{e,\mathrm{ped}}$}
    \end{subfigure}
    \caption{SHAP analyses illustrating feature importances for (a)~$n_{e,\mathrm{ped}}$ and (b)~$T_{e,\mathrm{ped}}$ in the H-mode all scenario.}
    \label{fig:pedestalfits_SHAP_diagrams}
\end{figure*}

In this appendix, we include the SHAP plots for $n_{e,\mathrm{ped} }$ and $T_{e,\mathrm{ped} }$ [\Cref{fig:pedestalfits_SHAP_diagrams}]. See \Cref{subsec:nepedTepedprediction} for more details.

\bibliographystyle{apsrev4-1}
\bibliography{EverythingPlasmaBib}

\end{document}